\documentclass[a4paper,11pt]{article}
\usepackage{jcappub, bm, color} 
\usepackage{amssymb,amsfonts,slashed,amsthm,amsmath,graphicx,soul}
\usepackage[caption=false]{subfig}
\usepackage{hyperref}
\usepackage{tikz,xcolor}
\usepackage{amsmath}
\usepackage{slashed}
\usepackage{comment}
\usepackage{cancel}
\usepackage{multirow}

\usepackage{slashed}

\renewcommand{\thefootnote}{\fnsymbol{footnote}}

\newcommand{\bea}{\begin{array}}
\newcommand{\eea}{\end{array}}
\newcommand{\beq}{\begin{eqnarray}}
\newcommand{\eeq}{\end{eqnarray}}

\newcommand{\eV}{ \ {\rm eV} }
\newcommand{\KeV}{ \ {\rm keV} }
\newcommand{\MeV}{\  {\rm MeV} }
\newcommand{\GeV}{\  {\rm GeV} }
\newcommand{\TeV}{\  {\rm TeV} }

\newcommand{\lmk}{\left(}  
\newcommand{\rmk}{\right)}
\newcommand{\lkk}{\left[}  
\newcommand{\rkk}{\right]}

\newcommand{\del}{\partial}  
\newcommand{\la}{\left\langle} 
\newcommand{\ra}{\right\rangle}

\newcommand{\Mpl}{M_{\rm Pl}}

\newcommand{\abs}[1]{\left\vert {#1} \right\vert}

\newcommand{\cphi}{\varphi}

\def\eq#1{Eq.~(\ref{#1})}

\newcommand{\DD}{\bar{D}}
\newcommand{\uu}{\bar{u}}
\newcommand{\dd}{\bar{d}}
\newcommand{\ee}{\bar{e}}

\definecolor{orange}{RGB}{255,100,0}
\definecolor{rosepink}{RGB}{248,100,100}

\begin{document}
\rightline{OU-HET-1133}
\rightline{RESCEU-2/22}
\rightline{TU-1146}

\vspace{-0.5cm}

\title{
Baryon Asymmetric Universe from Spontaneous CP Violation
}

\author{
Kohei Fujikura,$^{1}$\footnote{
E-mail address: fujikura@resceu.s.u-tokyo.ac.jp} \ 
Yuichiro Nakai,$^{2}$\footnote{
E-mail address: ynakai@sjtu.edu.cn} \ 
Ryosuke Sato,$^{3}$\footnote{
E-mail address: rsato@het.phys.sci.osaka-u.ac.jp} \\ 
Masaki Yamada$^{4,5}\footnote{
E-mail address: m.yamada@tohoku.ac.jp}$\\*[20pt]
$^1${\it \normalsize
Research Center for the Early Universe (RESCEU), Graduate School of Science,
The University of Tokyo, Hongo 7-3-1 Bunkyo-ku, Tokyo 113-0033, Japan} \\*[5pt]
$^2${\it \normalsize 
Tsung-Dao Lee Institute and School of Physics and Astronomy, \\ Shanghai Jiao Tong University, 800 Dongchuan Road,
Shanghai 200240, China} \\*[5pt]
$^3${\it \normalsize
Department of Physics, Osaka University, 
Toyonaka, Osaka 560-0043, Japan}  \\*[5pt]
$^4${\it \normalsize
Department of Physics, Tohoku University, 
Sendai, Miyagi 980-8578, Japan}  \\*[5pt]
$^5${\it \normalsize
Frontier Research Institute for Interdisciplinary Sciences, Tohoku University, \\
Sendai, Miyagi 980-8578, Japan}  \\*[50pt]
\vspace{-1.3cm}}

\abstract{
Spontaneous CP violation, such as the Nelson-Barr (NB) mechanism, is an attractive scenario
for addressing the strong CP problem 
while realizing the observed phase of the Cabibbo-Kobayashi-Maskawa (CKM) quark-mixing matrix. 
However, not only the CKM phase but also the baryon asymmetric Universe requires sources of CP violation. 
In this study, we show that 
a supersymmetric NB mechanism can naturally accommodate
the Affleck-Dine (AD) baryogenesis within a CP-invariant Lagrangian.
The model provides flat directions associated with new heavy quarks.
Focusing on one of the directions, we find that the correct baryon asymmetry is obtained with a sufficiently low reheating temperature which does not cause the gravitino problem.
Some parameter space is consistent with the gravitino dark matter.
We assess radiative corrections to the strong CP phase induced
by gauge-mediated supersymmetry breaking and CP-violating heavy fields
and show that the strong CP problem is solved in a viable parameter space
where the visible sector supersymmetric particles must be lighter than $\mathcal{O}(100) \, \rm TeV$.
Even in the case that they are heavier than the TeV scale,
our model predicts the neutron electric dipole moment
within the reach of the near future experiments.
Our model addresses the electroweak naturalness problem, strong CP problem, baryon asymmetric Universe,
and dark matter.
Then, the model may give a new compelling paradigm of physics beyond the Standard Model. 
}

\maketitle

\renewcommand{\thefootnote}{\arabic{footnote}}
\setcounter{footnote}{0}

\section{Introduction}

The electroweak naturalness and the strong CP problem are two major issues of fine-tuning in the Standard Model (SM).
The electroweak scale is unstable under quantum effects
and significant fine-tuning is required to set it to a much smaller scale than the UV energy scales, such as the Planck scale.
An elegant solution to this problem is provided by supersymmetry (SUSY):
quadratically divergent contributions to the Higgs mass-squared parameter are canceled by those of superpartners.
However, if SUSY is realized in nature, it must be spontaneously broken.
SUSY breaking introduces numerous new parameters into the theory.
Arbitrary choices of these parameters lead to dangerous CP and flavor violating processes,
and the superpartner mass must be heavier than about 1000 TeV (see, \textit{e.g.}, Ref.~\cite{UTfit:2007eik}), which reintroduces a need for fine-tuning.
This issue can be addressed using gauge-mediated SUSY breaking
(see Refs.~\cite{Giudice:1998bp,Kitano:2010fa} for reviews).
Superpartner masses in the visible sector are generated via flavor-independent SM gauge interactions. 
Gauge mediation predicts a light gravitino as the lightest supersymmetric particle (LSP).
To avoid overproduction of this light gravitino, the reheating temperature of the Universe must be sufficiently low \cite{Pagels:1981ke,Weinberg:1982zq,Khlopov:1984pf,Moroi:1993mb}. 

The strong CP problem is the question of why a large CP violation has never been observed in QCD~\cite{Jackiw:1976pf, Callan:1976je, Peccei:1977hh}.
The absence of observation of the neutron electric dipole moment (EDM) leads to a significant constraint on the effective $\theta$-angle, such as 
$\bar{\theta} \lesssim 10^{-10}$
\cite{Baker:2006ts,Pendlebury:2015lrz}.
The most popular solution to the strong CP problem is the Peccei-Quinn (PQ) mechanism
\cite{Peccei:1977hh}, where
the dynamical axion \cite{Weinberg:1977ma,Wilczek:1977pj}
associated with the spontaneous breaking of the $U(1)_{\rm PQ}$ global symmetry sets $\bar{\theta}$ to zero at the potential minimum.
However, it has been discussed that the quantum gravity effects explicitly break any global symmetry.
This introduces a fine-tuning of the theory~\cite{Georgi:1981pu, Dine:1986bg, Giddings:1988cx, Coleman:1988tj, Gilbert:1989nq, Barr:1992qq, Kamionkowski:1992mf, Holman:1992us, Ghigna:1992iv, Dobrescu:1996jp,Carpenter:2009zs,Banks:2010zn}.%
\footnote{
There are several models that allow for the quality of the $U(1)_{\rm PQ}$ symmetry
by accidental symmetry from discrete gauge symmetries~\cite{Chun:1992bn, BasteroGil:1997vn,Babu:2002ic,Dias:2002hz,Harigaya:2013vja,Harigaya:2015soa},
abelian gauge symmetries~\cite{Fukuda:2017ylt, Duerr:2017amf, Fukuda:2018oco, Bonnefoy:2018ibr, Ibe:2018hir},
and non-abelian gauge symmetries~\cite{Randall:1992ut,Dobrescu:1996jp,Redi:2016esr,DiLuzio:2017tjx,Lillard:2018fdt,Lee:2018yak,Gavela:2018paw,Buttazzo:2019mvl,Ardu:2020qmo, Yin:2020dfn}.
In Refs.~\cite{Cheng:2001ys,Izawa:2002qk,Fukunaga:2003sz,Choi:2003wr, Izawa:2004bi,Flacke:2006ad,Kawasaki:2015lea,Yamada:2015waa,Cox:2019rro,Yamada:2021uze,Lee:2021slp}, models with extra dimensions were also considered.
Furthermore, a model based on superconformal dynamics was considered in Ref.~\cite{Nakai:2021nyf}.
}
Moreover, depending on the scenarios of the $U(1)_{\rm PQ}$ symmetry breaking, there are challenges in the field of cosmology, including the domain wall problem~\cite{Sikivie:1982qv} and isocurvature problem~\cite{Axenides:1983hj,Seckel:1985tj,Linde:1985yf,Linde:1990yj,Turner:1990uz,Lyth:1991ub}.
Although the effects on cosmological history by the supersymmetric partners, axino and saxion, are highly model dependent
(see, e.g., Refs.~\cite{Kim:2008hd,Kawasaki:2013ae,Choi:2013lwa}), they could still cause a cosmological problem.
An alternative approach to the strong CP problem is to assume that the theory respects CP symmetry, but breaks it spontaneously 
to realize the observed Cabibbo-Kobayashi-Maskawa (CKM) phase.\footnote{
The idea of spontaneous CP violation has been used in SUSY to suppress contributions to EDMs
\cite{Nir:1996am,Aloni:2021wzk,Nakai:2021mha}.}
The Nelson-Barr (NB) mechanism \cite{Nelson:1983zb,Barr:1984qx,Barr:1984fh} introduces a vector-like heavy quark, whose realization does not regenerate a dangerous $\bar{\theta}$.
As the mechanism also requires new scalar fields to break the CP, which causes another naturalness problem,
a supersymmetric extension of the model is a natural possibility \cite{Dine:1993qm}. 
However, 
CP and flavor violations in SUSY breaking parameters are still significantly constrained
to avoid the regeneration of a sizable $\bar{\theta}$.
The constraints are stronger than those of flavor-changing neutral currents and are independent of superpartner mass scales.
Therefore, gauge-mediated SUSY breaking is 
necessary in this framework
\cite{Dine:2015jga,Evans:2020vil}.

Considering spontaneous CP violation, 
it should be noted that not only the CKM phase but also the baryon asymmetric Universe requires sources of CP violation. 
Therefore, it is natural to consider how the baryon asymmetric Universe
is realized if CP symmetry, whose violation is one of the necessary conditions to generate the baryon asymmetry,
is fundamentally preserved in the Lagrangian.
In addition, baryogenesis presents several challenges in the minimal supersymmetric Standard Model (MSSM).
Although thermal leptogenesis \cite{Fukugita:1986hr} is one of the most common scenarios for generating baryon asymmetry,
this requires a high reheating temperature, which leads to the overproduction of gravitinos in gauge mediation.
The Affleck-Dine (AD) mechanism \cite{Affleck:1984fy, Murayama:1993em, Dine:1995kz}
is an alternative method for baryogenesis that is compatible with a low reheating temperature in a supersymmetric theory.
The mechanism uses a flat direction, called an AD field, that carries a nonzero baryon or lepton number.
The AD field develops a large expectation value during inflation, 
after which a coherent oscillation is produced,
and a sizable baryon or lepton charge density is stored.
Finally, the condensate of the AD field decays, converting the charge density to a baryon asymmetry via the electroweak anomaly.
However, in the MSSM with gauge-mediated SUSY breaking,
there are challenges to this option too. 
The AD mechanism is preceded~\cite{Kusenko:1997si,Enqvist:1997si,Kasuya:1999wu,Kasuya:2000wx,Kasuya:2001hg} by the formation of non-topological solitons, called Q-balls~\cite{Coleman:1985ki,Dvali:1997qv,Kusenko:1997zq,Kusenko:1997ad}.
The Q-balls may or may not dominate the Universe and decay into quarks (baryons), gravitinos, and next-to-lightest SUSY particles.
In this scenario, the correct relic abundance of dark matter (DM) and the baryon asymmetry are obtained only in a limited parameter space~\cite{Kasuya:2001hg,Shoemaker:2009kg,Doddato:2011fz,Kasuya:2011ix,Kasuya:2012mh}.

In this study, we discuss a viable possibility for generating baryon asymmetry
in the framework of spontaneous CP violation based on the supersymmetric NB mechanism.%
\footnote{
Ref.~\cite{Evans:2020vil} assumes thermal leptogenesis in the supersymmetric NB model,
which may encounter the gravitino overproduction problem. 
In Ref.~\cite{Valenti:2021xjp}, the authors commented on the possibility of realizing resonant leptogenesis
in a non-SUSY model of spontaneous CP violation,
wherein a certain amount of fine-tuning is required to generate a sufficient amount of baryon (or $B-L$) asymmetry. 
}
The source of CP violation is the spontaneous CP violations that are driven by an AD field during inflation,
which are different from the source of spontaneous CP violation at present. 
Since the AD field is a complex scalar field, 
its large vacuum expectation value (VEV) violates the CP as well as the baryon (and/or lepton) symmetry.
The AD mechanism, therefore, works with the CP-conserving Lagrangian.
As mentioned earlier, the supersymmetric NB model contains a new vector-like heavy quark, and the flat directions associated with these quarks
can be used to generate baryon asymmetry via the AD mechanism.
Since the heavy quark has an explicit supersymmetric mass term, 
there is no danger of long-lived Q-ball formation.
Even if Q-balls are formed, they immediately decay into lighter SUSY particles.
Also, the reheating temperature can be low enough to avoid the gravitino overproduction problem.
In parameter regions that are consistent with the correct baryon asymmetry,
the gravitino LSP can give the correct DM abundance.
The model then simultaneously addresses the electroweak naturalness problem, strong CP problem, baryon asymmetry,
and cosmological DM.
The logic leading to the current scenario is summarized in Table~\ref{tab1},
which shows that this is a natural direction for solving fine-tuning problems as well as cosmological issues.
Interestingly, we find that our model predicts the neutron EDM within the reach of the near future experiments.

\begin{table*}[t]
\begin{center}
\scalebox{0.89}[0.9]{
\begin{tabular}{c||c|c|c}
	\rule[-5pt]{0pt}{20pt} Problems & Solutions & Challenges & NB with gauge mediation?  \\
	\hline	\hline
	\rule[-7pt]{0pt}{20pt} Hierarchy & SUSY & Gauge mediation is favored & $\checkmark$ 
	\\
	\hline
	\multirow{3}{*}{Strong CP}
	 &\multirow{2}{*}{PQ mechanism} & 
	 \multirow{2}{*}{\hspace{-0.5cm}\Bigg\{}
	 \rule[-7pt]{0pt}{20pt}  \hspace{-0.4cm}Quality problem& 
	\\
	 &  &  \rule[-7pt]{0pt}{20pt}\hspace{-0.5cm}~~~~~~ Cosmological issues & 
	\\
	& \rule[-7pt]{0pt}{20pt} SCPV (NB) & Gauge mediation is necessary & $\checkmark$ 
	\\
	\hline
	\multirow{2}{*}{BAU} & \rule[-7pt]{0pt}{20pt} Thermal LG & Gravitino problem in SUSY & No CP-violating source
	\\
	 & \rule[-7 pt]{0 pt}{20 pt} ADBG & Maybe too many Q-balls in the MSSM & $\checkmark$ (The present work)
\end{tabular}
\label{tab1}}
\end{center}
\caption{Summary of possible solutions for the problems in the SM and their challenges in 3+1 dimensions
(NB = the NB model; SCPV = spontaneous CP violation; BAU = baryon asymmetric Universe; LG = leptogenesis; ADBG = Affleck-Dine baryogenesis).}
\end{table*}

To confirm that the supersymmetric NB model gives a viable solution to the strong CP problem,
it is important to assess radiative corrections to $\bar \theta$.
Such radiative corrections have been investigated in NB models without SUSY~\cite{Bento:1991ez}
and with SUSY in gravity mediation~\cite{Dine:1993qm,Barr:1993hb}.
Refs.~\cite{Dine:2015jga,Evans:2020vil} initiated to study radiative corrections to $\bar{\theta}$
in the supersymmetric NB model with gauge-mediated SUSY breaking
while a detailed analysis has not been done so far.
We then provide a further study of this issue.
When the messenger scale of gauge mediation is lower than the scale of spontaneous CP violation,
the radiative corrections can be analyzed by using an effective field theory (EFT) consisting of the MSSM and messenger fields.
In this paper, we explicitly construct such an EFT by integrating out new fields in the NB model
and estimate the radiative corrections.
We find that the corrections to $\bar{\theta}$ are suppressed by the ratio of the messenger scale
to the scale of SCPV.
Our result smoothly connects with the case that the messenger scale is higher than the SCPV scale.

The rest of the paper is organized as follows.
In Sec.~\ref{sec:model}, we review the supersymmetric NB mechanism to obtain a solution to the strong CP problem
and analyze the observed CKM phase.
Sec.~\ref{radiative} includes a detailed discussion of radiative corrections to $\bar{\theta}$ 
induced by gauge-mediated SUSY breaking and CP-violating heavy fields.
Sec.~\ref{sec:ADBG} discusses the AD mechanism for baryogenesis in the model.
Focusing on one of the flat directions associated with new heavy quarks,
we find that the correct baryon asymmetry is obtained
with a sufficiently low reheating temperature to avoid the gravitino problem.
In Sec.~\ref{plots}, we determine the parameter space in which the strong CP problem, baryon asymmetric Universe,
and DM are explained.
Sec.~\ref{sec:conclusion} is devoted to conclusions and discussions.

\section{SUSY Nelson-Barr model} \label{sec:model}
We introduce a vector-like pair of heavy down-type quark chiral superfields $D$ and $\DD$,
and the SM singlet chiral superfields $\eta_\alpha$ ($\alpha = 1 , \ldots , n_\eta$),
whose scalar components develop complex VEVs and break CP symmetry spontaneously. 
To accommodate a physical CP-breaking phase, $n_\eta \ge 2$ is required.
In this paper, we do not specify the $\eta$ sector and just assume they have VEVs. 
The representations of these fields under the SM gauge symmetry $SU(3)_C\times SU(2)_W\times U(1)_Y$ are summarized as follows:
\begin{align}
D : \left({\bf 3}, {\bf 1}, -\frac{1}{3} \right), \qquad \bar{D}:\left({\bf \bar{3}}, {\bf 1}, +\frac{1}{3}\right),
\qquad\eta_\alpha :({\bf 1}, {\bf 1}, 0).
\end{align}
We assume the model respects a discrete $Z_N$ symmetry,
\begin{align}
	D \rightarrow e^{2\pi i/N}D,\quad \bar{D}\rightarrow e^{-2\pi i /N}\bar{D},\quad \eta_\alpha \rightarrow e^{-2\pi i/N}\eta_\alpha \,. \label{eq:Z2 symmetry}
\end{align}
Then, the superpotential of the model is given by
\begin{align}
	W_{\rm NB}=y^D_{\alpha i}\eta_\alpha D \bar{d}_i +M_D D\bar{D}+W_{\rm MSSM} \, , \label{eq:superpotential NB}
\end{align}
where we sum over repeated indices,
$\bar{d}_i$ ($i=1,2,3$) denote three generations of MSSM right-handed down-type quark supermultiplets,
$y^D_{\alpha i}$ are Yukawa couplings, $M_D$ is a mass parameter, and
$W_{\rm MSSM}$ denotes the superpotential of the MSSM, which includes
\begin{align}
	W_{\rm MSSM} \supset - y^u_{ij} \bar{u}_i Q_j H_u+y^d_{ij} \bar{d}_i Q_j H_d +\mu H_u H_d \, .\label{eq:superpotential MSSM}
\end{align}
In this case, $Q_j, \bar{u}_i$, and $H_{u,d}$ are the MSSM left-handed quark, the right-handed up-type quark,
and the up and down-type Higgs supermultiplets, respectively.
As leptons do not play a role in this section, we have omitted the interactions with leptons.
Other renormalizable operators such as $Q\bar{D}H_d$ and $\eta_\alpha D\bar{D}$ are forbidden by $Z_N$ symmetry.
Given that CP symmetry is assumed to be exact in the NB model, all parameters in the superpotential interactions \eqref{eq:superpotential NB} and \eqref{eq:superpotential MSSM} are real.

After the scalar component of $\eta_\alpha$ develops a VEV with a non-vanishing CP-violating phase,
we have obtained the following $4\times 4$ down-type quark mass matrix:
\begin{align}
\mathcal{L}_{\rm mass} =  \bigl(\bar{d}\,~\bar{D}\bigr)\mathcal{M}
\Biggl( \begin{matrix}
     d \\ D
\end{matrix}  \Biggr) +{\rm h.c.} \, , \qquad \mathcal{M} =
	\begin{pmatrix}
		m_d & B\\
		0 & M_D
	\end{pmatrix},
\end{align}
where $(B)_i \equiv y^D_{\alpha i} \langle \eta_\alpha \rangle$ is a $3\times 1$ matrix containing a nonzero CP phase.
The MSSM down-type quark mass $m_d \equiv y^d \langle H_d \rangle$ is assumed to be much smaller than $M_D$.
Based on the structure of $\mathcal{M}$, we can find that ${\rm arg}({\rm det}\mathcal{M})=0$,
and thus, the strong CP phase $\bar{\theta}$ does not appear at the tree level.
Let us now decompose $\bar{d}, \bar{D}, d, D$
into light ($\hat{\bar{d}}, \hat{d}$) and heavy ($\hat{\bar{D}}, \hat{D}$) states, which are defined as
\begin{align}
\bigl(\bar{d}\,~\bar{D}\bigr)
	 =
	\bigl(\hat{\bar{d}}\,~\hat{\bar{D}}\bigr)U_R^\dag\,, \qquad 
	\Biggl( \begin{matrix}
     d \\ D
\end{matrix}  \Biggr)
	 =
	U_L \Biggl( \begin{matrix}
    \hat{d} \\ \hat{D}
\end{matrix}  \Biggr), \label{eq:unitary transformations}
\end{align}
with the $4\times 4$ unitary matrices $U_R= (u_i, u_4)$, $U_L=(v_i, v_4)$,
to diagonalize $\mathcal{M}\mathcal{M}^\dag$ and $\mathcal{M}^\dag \mathcal{M}$
(neglecting the small mass $m_d$), respectively,
where $u_i, v_i$ are the eigenvectors of the light SM quarks, and $u_4, v_4$ are those of the heavy quark.
The eigenvectors of the heavy quarks are explicitly given by
\begin{align}
	u_4=\frac{1}{M_{\rm CP}}
	\Biggl( \begin{matrix}
    B \\ M_D
\end{matrix}  \Biggr),
\qquad v_4 =\Biggl( \begin{matrix}
    0 \\ 1
\end{matrix}  \Biggr), \label{eq:eigenvector}
\end{align}
where $M_{\rm CP}^2 \equiv M_D^2 + B^\dag B$ is the physical mass squared of the heavy quark.
In the following discussion, we take the flavor basis $(v_i)_j=\delta_{ij}$, where the mass matrix of the light states $\hat{\bar{d}}, \hat{d}$ is given by $(m_{\hat{d}})_{ij}=(u^*_i)_k (m_d)_{kj}$.
Considering the up-type quark for which mass matrix is diagonal, the CKM matrix $K$ is defined by the matrix that
diagonalizes $m_{\hat{d}}^\dag m_{\hat{d}}$.
Hence, we find that $\overline{m}_d^2=K^\dag m_d^Tu_iu_i^\dag m_d K$, where $\overline{m}_d \equiv {\rm diag} (m_d,m_s,m_b)$ is the diagonal SM down-type quark mass matrix.
Using $U_R U_R^\dag ={\bm 1}_{4\times 4}$, we can obtain that
\begin{align}
	u_iu_i^\dag +u_4u_4^\dag={{\bm 1}}_{4\times 4} \, . \label{eq:projection}
\end{align}
This equation and Eq.~\eqref{eq:eigenvector}
lead to the following relation \cite{Bento:1991ez,Cherchiglia:2020kut,Cherchiglia:2021vhe,Valenti:2021rdu}:
\begin{align}
 K \overline{m}_d^2 K^{\dag} \simeq m_d^T m_d - \dfrac{m_d^T B B^\dag m_d}{M_{\rm CP}^2} \, .	
\end{align}
Thus, it can be found that a CP-violating phase vanishes for $M_D^2\gg B^\dag B$.
Then, we assume $M_D^2\sim B^\dag B$ in the following discussion.

The SM contribution to $\bar{\theta}$ originating from the CKM phase 
is much smaller than $10^{-10}$ \cite{Dugan:1984qf,Ellis:1978hq,Khriplovich:1985jr} because of the significant suppression by the Jarlskog invariant and the GIM mechanism. 
To address the strong CP problem, we must also make sure that radiative corrections and 
Planck-suppressed operators
do not generate a large strong CP phase.
We will discuss radiative corrections to $\bar{\theta}$ induced
by gauge-mediated SUSY breaking and CP-violating heavy fields in the next section.
Here, our focus is on the effect of the Planck-suppressed operators.
For concreteness, we define the effective strong CP phase that is invariant by field rotations as: 
\begin{align}
	\bar{\theta} \equiv \theta-{\rm arg}\,{\rm det}(y^u y^{d}) -3\,{\rm arg}\,(m_{\tilde{g}}) \, , \label{eq:theta bar}
\end{align}
where $m_{\tilde{g}}$ is the gluino mass. 
Corrections to the phase of the quark and gluino masses
must be suppressed by a factor of $10^{-10}$ to solve the strong CP problem.
The lowest order Planck-suppressed operators in the superpotential 
are given by
\begin{align}
W_{\rm CP} = \frac{c_{1,\alpha}}{M_{\rm Pl}^{N-1}} \eta_\alpha^{N} H_u H_d
+ \frac{c_{2,\alpha}}{M_{\rm Pl}^{N-1}} \eta_\alpha^N D\DD
+ \frac{c_{3,\alpha}}{M_{\rm Pl}^{N-1}} \eta_\alpha^{N-1} Q \bar{D} H_d \, ,
\label{plancksuppressed}
\end{align}
where $c_{1,\alpha}, c_{2,\alpha}, c_{3,\alpha}$ are $\mathcal{O}(1)$ dimensionless constants, and $\Mpl$ ($\simeq 2.4 \times 10^{18} \GeV$) is the reduced Planck scale.
The first term generates a CP-violating $\mu$-term after $\eta_\alpha$ develops a complex VEV.
This CP-violating $\mu$-term leads to a correction of the phase of the quark masses
through a loop of the Higgsino with the insertion of the $\mu$ and $A$-terms.
It has been shown that $\bar{\theta} < 10^{-10}$ requires ${\rm arg}\,(\mu) < 10^{-8} \tan\beta$~\cite{Hiller:2002um}, 
which turns out to be 
\begin{align}
	\langle \eta_\alpha \rangle \lesssim 10^{-8/N} \left( \frac{{\rm Re}\, (\mu) \tan\beta}{M_{\rm Pl}}\right)^{1/N}M_{\rm Pl} \, , \label{eq:CP-violating mu-term}
\end{align}
for 
${\rm Re}(\mu)\gg {\rm Im}\, (\mu)$. 
The second and third terms in Eqs.~\eqref{plancksuppressed} may lead to a CP-violating phase of the order:
$(\langle \eta_\alpha \rangle /M_{\rm Pl})^{N-1}$.
This phase must be smaller than $10^{-10}$, which requires $\langle \eta_\alpha \rangle <10^{-10/(N-1)}M_{\rm Pl} $.
There can also be Planck-suppressed K\"ahler potential terms such as 
\begin{align}
K_{\rm CP} \supset \frac{c_{\alpha \beta,ij}}{M_{\rm Pl}^2}
\eta_\alpha^\dagger \eta_\beta \bar{d}^\dagger_i \bar{d}_j +{\rm h.c.} ,
\label{eq:Kahler contribution}
\end{align}
with $\mathcal{O}(1)$ dimensionless coefficients $c_{\alpha \beta,ij}$. 
This term generates non-canonical CP-violating kinetic terms for the $\bar{d}_i$ fields.
However, it has been noted in Refs.~\cite{Hiller:2001qg,Hiller:2002um} that, owing to the hermiticity of the wavefunction renormalization factors, the kinetic terms are canonically normalized
without an additional contribution to the strong CP phase.
Hence, the K\"ahler potential terms, such as those in Eq.~\eqref{eq:Kahler contribution} do not lead to an additional
constraint on the model parameters.

To set the cosmological constant to zero, we require $\langle W \rangle \sim m_{3/2}M_{\rm Pl}^2$
where $m_{3/2}$ is the gravitino mass.
This term generally has an $\mathcal{O}(1)$ complex phase which
generates a gluino mass phase via the anomaly mediation at the one-loop order \cite{Randall:1998uk, Giudice:1998xp}.
Thus, the following constraint can be found
\cite{Dine:2015jga}:
\begin{align}
\frac{\alpha_s}{4\pi}  \frac{m_{3/2}}{m_{\tilde{g}}}	< 10^{-10}. \label{eq:anomaly mediation}
\end{align}
In the present work, we focus on gauge-mediated SUSY breaking, wherein $m_{3/2}$ is much smaller
than the gluino mass $m_{\tilde{g}}$.
Hence, this constraint can be fulfilled by having a sufficiently low value for the messenger mass scale.

\section{Radiative corrections}
\label{radiative}

We now discuss radiative corrections to $\bar{\theta}$ induced by the soft SUSY breaking parameters and CP-violating heavy fields.
Radiative corrections to $\bar{\theta}$ in NB models have been investigated without SUSY~\cite{Bento:1991ez}
and with SUSY in gravity mediation~\cite{Dine:1993qm,Barr:1993hb}.
In Ref.~\cite{Dine:2015jga}, it was argued that radiative corrections to $\bar{\theta}$ are quite problematic in those cases,
whereas they are controlled in gauge-mediated SUSY breaking, as we will see below.

In the exact supersymmetric theory, the superpotential is not renormalized
because of the non-renormalization theorem~\cite{Ellis:1982tk}.
In addition, $\bar{\theta}$ is not renormalized by the wavefunction renormalization as discussed in the previous section.\footnote{
Even if we include threshold corrections to $\bar{\theta}$ that is generated by integrating heavy fields,
this conclusion does not change provided that the theory is exactly supersymmetric.
See Ref.~\cite{Hiller:2002um} for detailed discussions.}
However, $\bar{\theta}$ is not protected against SUSY breaking.
Hence, radiative corrections to $\bar{\theta}$ depend on the SUSY breaking scale.
The mass scale at which SUSY breaking effects are mediated to the visible sector is denoted by $M_{*}$.
It $M_{*}$ is given by a messenger mass scale in gauge mediation and by the Planck scale in gravity mediation.
In the case of $M_{\rm CP} > M_*$, we can analyze a correction to the strong CP phase $\delta\bar\theta$
by using an EFT with the MSSM fields and messengers.
Since there is no SUSY breaking source at the scale of $M_{\rm CP}$,
this EFT is a supersymmetric theory with higher dimensional operators suppressed by $1/M_{\rm CP}$.
The radiative correction to $\bar{\theta}$ in the EFT should involve both SUSY breaking and one of the higher dimensional operators, and we can see a suppression factor $1/M_{\rm CP}^2$ in $\delta\bar\theta$.
Although this analysis is valid only if $M_* < M_{\rm CP}$, there exists a contribution to $\delta\bar\theta$ in the case of $M_* > M_{\rm CP}$ which is smoothly connected at $M_* \sim M_{\rm CP}$.
In this section, we mostly focus on the EFT analysis on $\delta\bar\theta$ for the case with $M_* < M_{\rm CP}$
and then discuss the case with $M_* > M_{\rm CP}$ briefly.

After the scalar components of $\eta_\alpha$ develop CP-violating VEVs,
the superpotential \eqref{eq:superpotential NB} is rewritten in the basis of $(\hat{\bar{d}},\hat{d},\hat{\bar{D}},\hat{D})$ as 
\begin{align}
	W_{\rm NB}= y^{\hat{d}}_{ij}\hat{\bar{d}}_iQ_j H_d+y^{Q\hat{\bar{D}}}_i \hat{\bar{D}}Q_i H_d +y^{\hat{D}}_{\alpha i}\delta \eta_\alpha \hat{D}\hat{\bar{d}}_i + y^{\hat{D}\hat{\bar{D}}}_\alpha \delta \eta_\alpha \hat{D}\hat{\bar{D}}+M_{\rm CP}\hat{D}\hat{\bar{D}} \, ,  \label{eq:superpotential mass basis}
\end{align}
where we have used Eqs.~\eqref{eq:unitary transformations}, \eqref{eq:eigenvector} and
$\delta \eta_\alpha\equiv \eta_\alpha - \langle \eta_\alpha \rangle$ parameterizes the fluctuations around the CP-breaking vacuum.
The coupling constants $y^{\hat{d}}_{ij},y^{Q\hat{\bar{D}}}_i,y^{\hat{D}}_{\alpha i},y^{\hat{D}\hat{\bar{D}}}_\alpha$
in the superpotential are defined by
\begin{align}
	y^{\hat{d}}_{ij}\equiv (u^*_i)_k y^d_{kj},
	\qquad y^{Q\hat{\bar{D}}}_i \equiv  (u^*_4)_ky^d_{ki},
	\qquad y^{\hat{D}}_{\alpha i}\equiv (u_i^*)_ky^D_{\alpha k},
	\qquad y^{\hat{D}\hat{\bar{D}}}_\alpha \equiv (u_4^*)_ky^D_{\alpha k} \, .
\end{align}
Let us now construct an effective theory below $M_{\rm CP}$ by integrating the heavy fields.
The $F$-term conditions, $\partial W_{\rm NB}/\partial \hat{\bar{D}}=\partial W_{\rm NB}/\partial \hat{D}=0$,
and $\delta \eta_\alpha =0$ lead to
\begin{align}
	\hat{\bar{D}}=0, \qquad \hat{D}=-\dfrac{y^{Q\hat{\bar{D}}}_i}{M_{\rm CP}}Q_iH_d  \,.
\end{align}
Then, the following effective K\"ahler potential is obtained at the tree level,
\begin{align}
	\Delta K_{\rm tree} = \dfrac{(y^{Q\hat{\bar{D}}}_j)^*y^{Q\hat{\bar{D}}}_i}{M_{\rm CP}^2}
	 (Q_j H_d)^\dagger Q_i H_d \label{eq:tree Kahler} \, .
\end{align}
One-loop corrections to the effective K\"ahler potential are expressed as
\cite{Intriligator:2007cp}
\begin{align}
	\Delta K_{\rm 1-loop} = -{\rm Tr}\,\left[\frac{M^\dag M}{32\pi^2}\log \left( \dfrac{M^\dag M}{\Lambda^2}\right)\right], 
	\qquad (M)_{ab}\equiv \frac{\partial^2 W_{\rm NB}}{\partial \Phi_a \partial \Phi_b} \, , 
	\label{eq:loop Kahler}
\end{align}
where $\Lambda \sim M_{\rm CP}$ is a renormalization scale and
$\Phi_a\equiv (\hat{D},\hat{\bar{D}},\delta \eta_\alpha)$.
Explicitly, the mass matrix $M$ is given by
\begin{align}
	M=
	\begin{pmatrix}
    0&M_{\rm CP}&y^{\hat{D}}\hat{\bar{d}} \,\\
    M_{\rm CP}&0&0\\
    y^{\hat{D}}\hat{\bar{d}} &0&M_\eta
    \end{pmatrix} . \label{eq:matrix}
\end{align}
where $M_\eta$ is the $n_\eta \times n_\eta$ mass matrix of the $\eta_\alpha$ field.
Assuming $M_{\rm CP} \ \sim M_ \eta$, we can expand the one-loop effective K\"ahler potential
with respect to $M_{\rm CP}^{-1}$ as given below.
\begin{equation}
	\Delta K_{\rm 1-loop} \sim
		-\dfrac{1}{32\pi^2}(y^{\hat{D}}_{\alpha i}\hat{\bar{d}}_i)^\dag y^{\hat{D}}_{\alpha j}{\hat{\bar{d}}}_j
		- \dfrac{1}{32\pi^2 M_{\rm CP}^2} (y^{\hat{D}}_{\alpha i}\hat{\bar{d}}_i)^\dag y^{\hat{D}}_{\alpha k}\hat{\bar{d}}_k
		( y^{\hat{D}}_{\beta j}\hat{\bar{d}}_j)^\dag y^{\hat{D}}_{\beta l}\hat{\bar{d}}_l + \mathcal{O}\left(\frac{1}{M_{\rm CP}^4}\right). \label{eq:one-loop Kahler}
\end{equation}
In this equation, the first term gives the non-canonically normalized kinetic term for $\hat{\bar{d}}$.
As discussed in the previous section, the wavefunction renormalization does not reintroduce $\bar{\theta}$.

\begin{figure}[t]
\begin{center}
\includegraphics[clip, width=4cm]{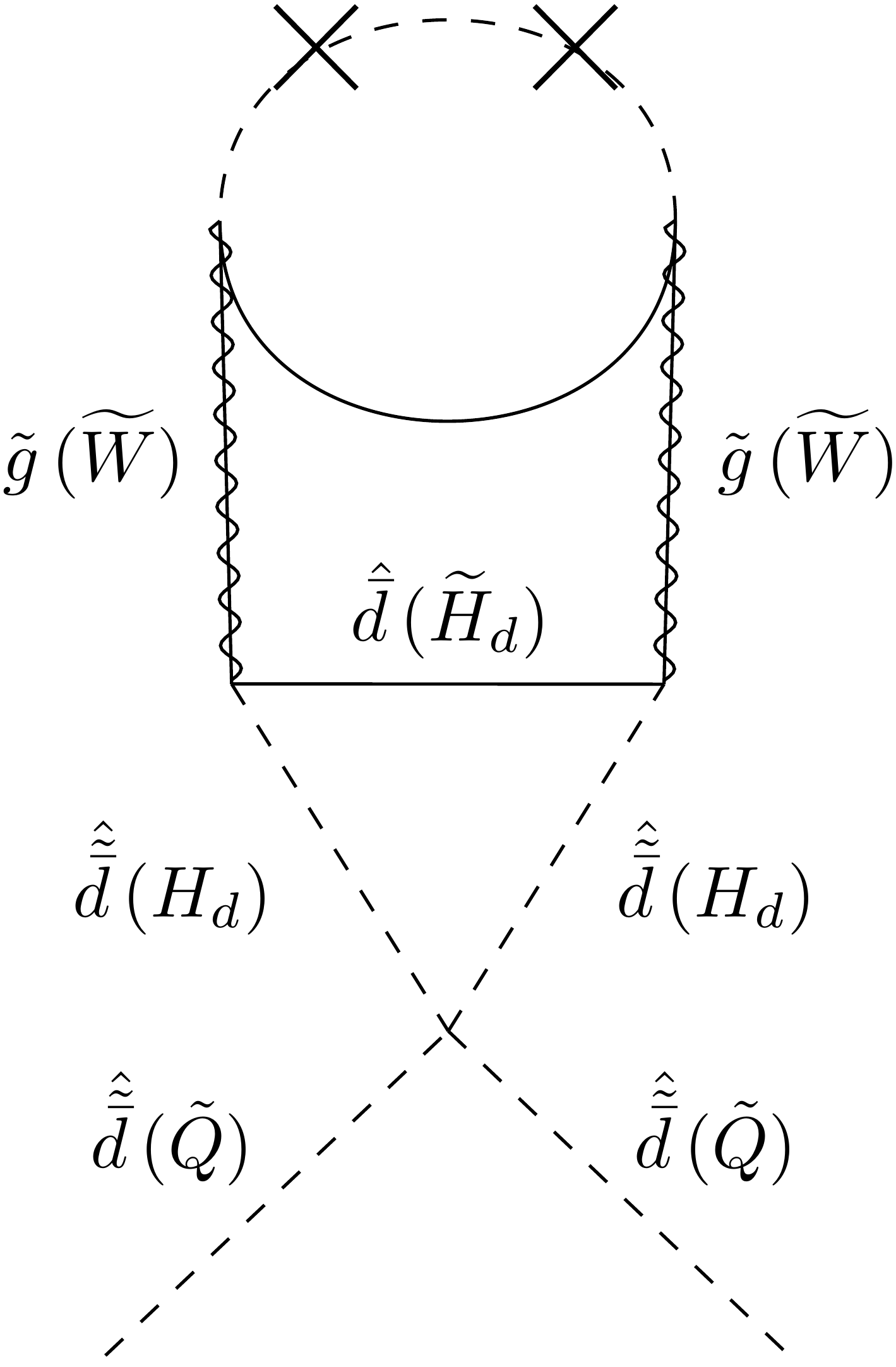}
\hspace{2cm}
\includegraphics[clip, width=3.6cm]{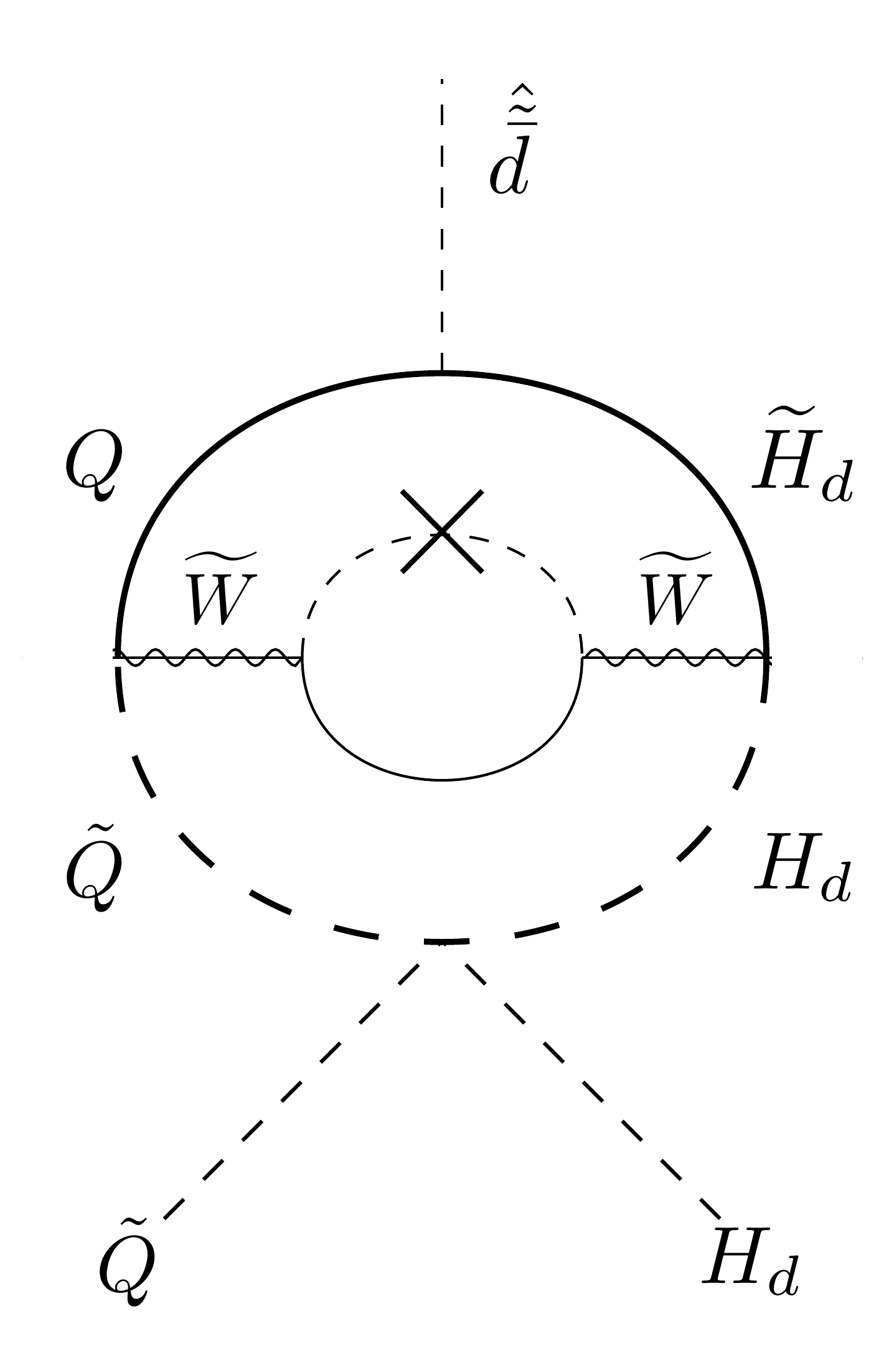}
\end{center}
\caption{The soft mass-squared parameters $m_{\hat{\tilde{\bar{d}}},\tilde{Q}}^2$ (left)
and the $A$-parameter $A^{ \hat{d}}$ (right) induced by the effective K\"ahler potential
obtained in Eqs.~\eqref{eq:tree Kahler}, \eqref{eq:loop Kahler}.
The cross denotes the insertion of the $F$-term of a SUSY breaking field.
\label{fig:SUSY breaking parameters}
}
\end{figure}

The relevant soft SUSY breaking terms generated by gauge mediation are
\begin{align}
	-\mathcal{L}_{\rm soft}\supset& \,\, \tilde{Q}_i^\dag (m_{\tilde{Q}}^2)_{ij} \tilde{Q}_j +\tilde{\bar{u}}_i (m_{\tilde{\bar{u}}}^2)_{ij}\tilde{\bar{u}}^\dag_j + \hat{\tilde{\bar{d}}}_i (m_{\hat{\tilde{\bar{d}}}}^2)_{ij} \hat{\tilde{\bar{d}}}_j^\dag \nonumber \\
	&+\left(\frac{1}{2}m_{\tilde{g}} \tilde{g}\tilde{g} -A^{u}_{ij} H_u \tilde{\bar{u}}_i\tilde{Q}_{j} +A^{\hat{d}}_{ij} H_d \hat{\tilde{\bar{d}}}_i  \tilde{Q}_j+{\rm h.c.}\right), \label{eq:SUSY breaking}
\end{align}
where $\tilde{X}$ denotes the superpartner of a SM field $X$.
In gauge mediation, the gluino mass is generated in the one-loop order:
$m_{\tilde{g}}=\alpha_s F/(4\pi M_*)$, where $\sqrt{F}$ is the SUSY breaking scale.
The soft scalar mass-squared parameters and $A$-parameters are generated at the two-loop order:
$m_{\rm soft}^2\sim \alpha^2 F^2/(16\pi^2 M_*^2)$
and $A^{u,\hat{d}}_{ij}\sim  y^{u,\hat{d}}_{ij} \alpha^2 F/ (16\pi^2 M_*)$
with $\alpha \equiv g^2/4\pi$ ($g$ denotes a SM gauge coupling).
A remarkable feature of gauge-mediated SUSY breaking is that
$m_{\tilde{Q}}^2,m_{\tilde{\bar{u}}}^2,m_{\hat{\tilde{\bar{d}}}}^2$ are flavor-universal
and $A^u,A^{\hat{d}}$ are proportional to the corresponding Yukawa couplings.
However, the effective K\"ahler potential obtained from Eqs.~\eqref{eq:tree Kahler}, \eqref{eq:loop Kahler},
leads to corrections to those soft SUSY breaking parameters
based on the diagrams shown in Fig.~\ref{fig:SUSY breaking parameters}.
These corrections are roughly evaluated as
\begin{align}
	&(\delta m_{\hat{\tilde{\bar{d}}}}^2)_{ij}
	\sim \left(\frac{1}{16\pi^2}\right)^2 (y^{\hat{D}}_{\alpha j} y^{\hat{D}}_{\beta k})^*y^{\hat{D}}_{\alpha k}y^{\hat{D}}_{\beta i} \left(\frac{M_*}{M_{\rm CP}} \right)^2 m_{\rm soft}^2 \, ,\nonumber\\[1ex]
	&(\delta m_{\tilde{Q}}^2)_{ij}
	\sim \frac{1}{16\pi^2}(y^{Q\hat{D}}_i)^* y^{Q\hat{D}}_j \left(\frac{M_*}{M_{\rm CP}} \right)^2 m_{\rm soft}^2 \, ,  \label{eq:flavor nonuniversal soft mass}\\[1ex]
	&(\delta A^{\hat{d}})_{ij}
	\sim 
	\frac{1}{16\pi^2} (y^{Q\hat{D}}_k)^* y^{Q\hat{D}}_j \left(\frac{M_*}{M_{\rm CP}} \right)^2
	A^{\hat{d}}_{ik} \nonumber \, .
\end{align}
Therefore, now the soft mass-squared parameters are not flavor-universal and
the $A$-parameters are not precisely proportional to the corresponding Yukawa couplings 
in the presence of the heavy fields.

\begin{figure}[t]
\begin{center}
\includegraphics[clip, width=7cm]{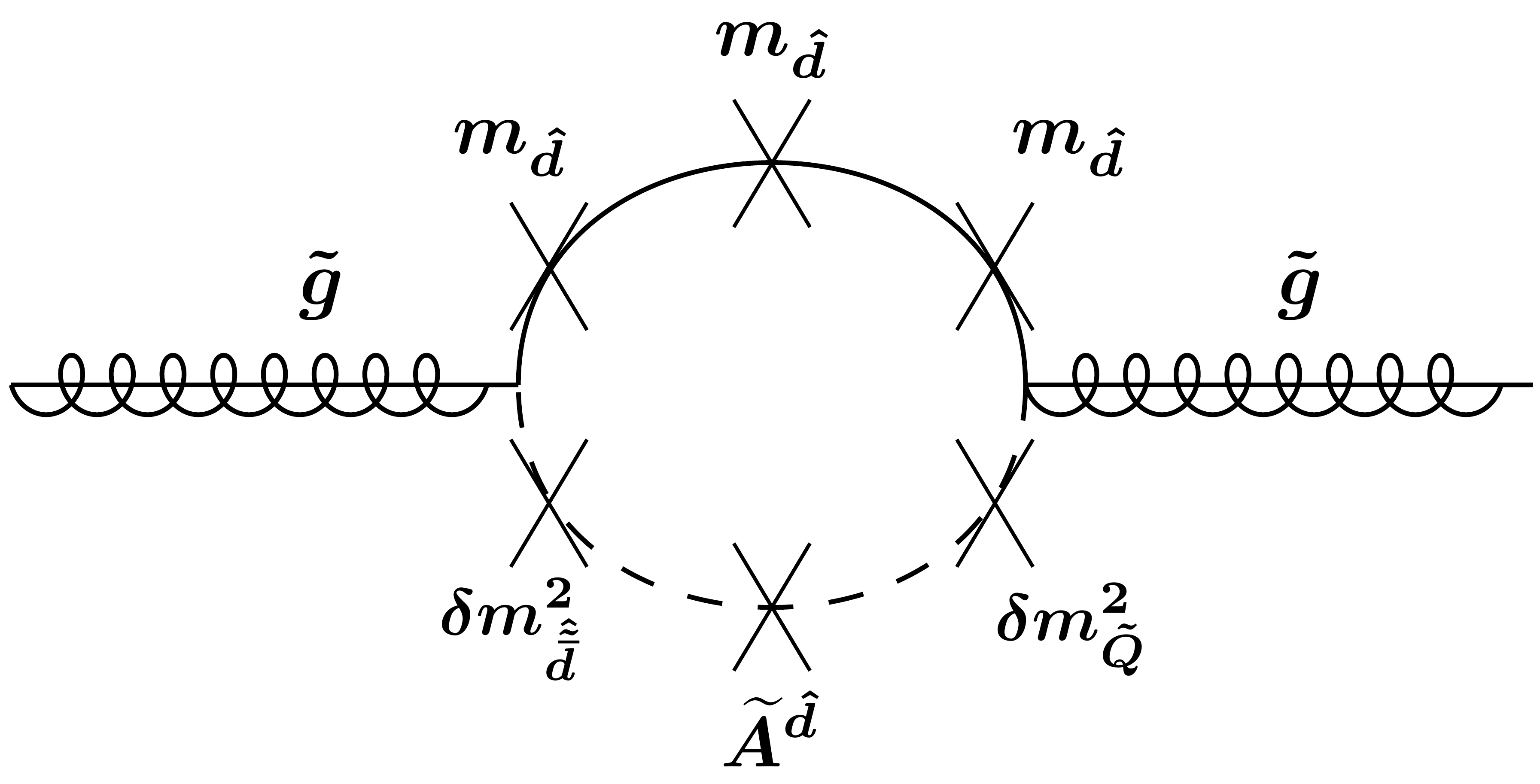}
\hspace{1cm}
\includegraphics[clip, width=7cm]{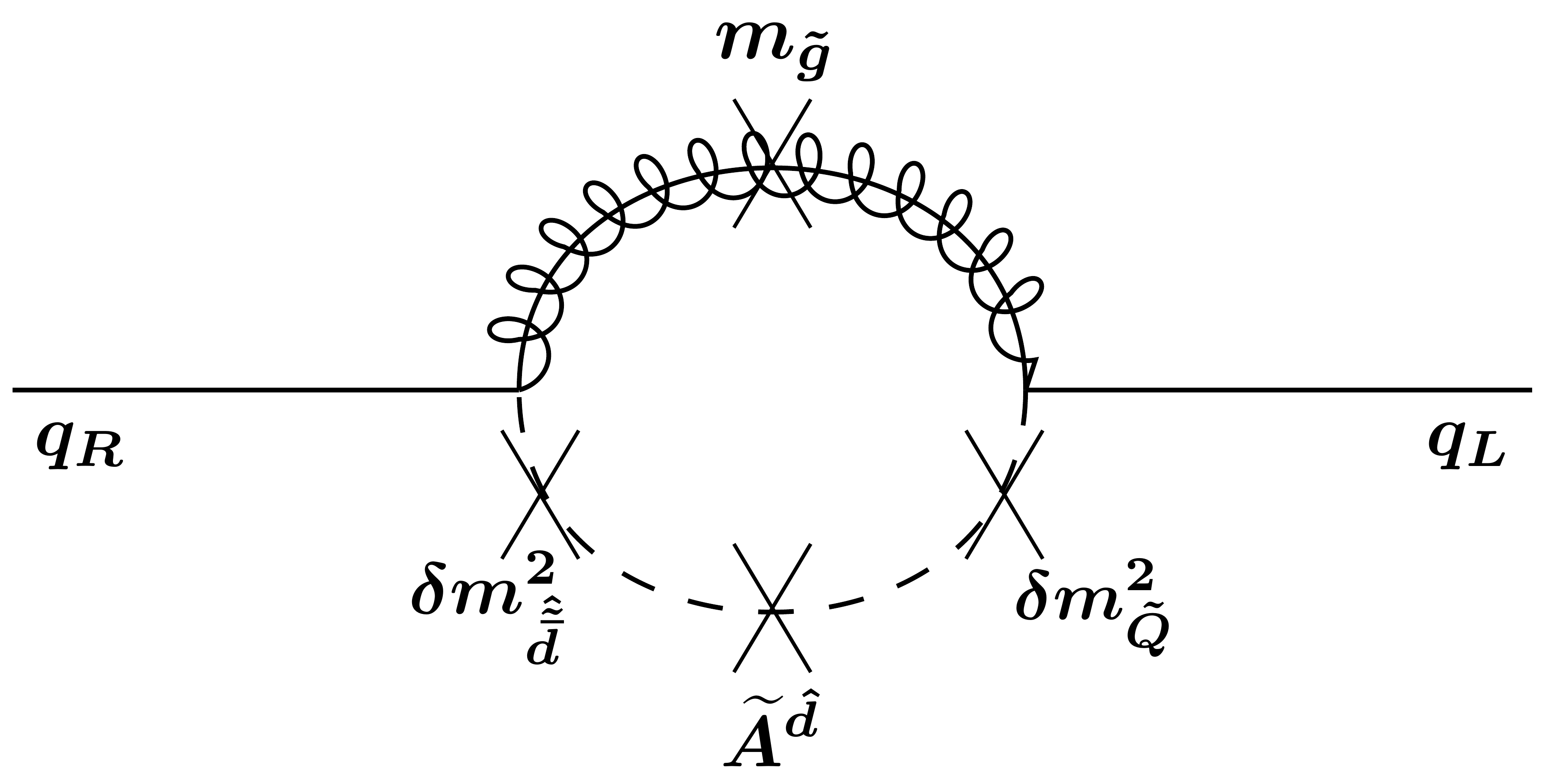}
\end{center}
\caption{ Radiative corrections to the phase of the gluino mass (the left panel) and quark mass (the right panel) induced by flavor violating SUSY-breaking parameters, $\delta m^2_{\hat{\tilde{\bar{d}}}}$ and $\delta m^2_{\tilde{Q}}$.
\label{fig:radiative corrections}
}
\end{figure}

The presence of soft SUSY breaking parameters~\eqref{eq:flavor nonuniversal soft mass}
induces complex phases to the SM quark and gluino masses, potentially leading to a large correction to $\bar{\theta}$.
Such a correction can be described as~\cite{Dine:1993qm}
\begin{align}
	\delta \bar{\theta} = -{\rm Im}\,{\rm Tr} \left( m_{\hat{d}}^{-1} \delta m_{\hat{d}} + m_u^{-1} \delta  m_u \right)- 3\, {\rm Im} \,(m_{\tilde{g}}^{-1} \delta m_{\tilde{g}} ) \, ,
\end{align}
where $\delta m_{\hat{d},u,\tilde{g}}$ denotes mass corrections of the $\hat{d},u,\tilde{g}$ fields
induced by the soft SUSY breaking parameters~\eqref{eq:flavor nonuniversal soft mass}.
To calculate the mass corrections, we treat the flavor non-universal parts of the soft mass-squared parameters,
$\delta m_{\hat{\tilde{\bar{d}}},\tilde{Q}}^2$, as perturbations.
The Feynman diagram of the lowest non-vanishing contribution to $\delta m_{\tilde{g}}$ with the CP-violating phase is shown in the left panel of Fig.~\ref{fig:radiative corrections}.
This contribution can be evaluated as~\cite{Hiller:2002um}\footnote{ The contribution to $\delta \theta^m_{\tilde{g}}$ from a single insertion of $m_{\hat{d}}$ vanishes because of the hermiticity of the matrices $\delta m_{\tilde{\hat{\bar{d}}}}^2$ and $\delta m_{\tilde{Q}}^2$. Therefore, the non-vanishing leading contribution arises from the three insertions of $m_{\hat{d}}$.}
\begin{equation}
	\begin{split}
	\delta \theta^{m}_{\tilde{g}} &\sim  \frac{\alpha_s}{4\pi}\dfrac{v_{d}^4}{m_{\tilde{g}}m_{\rm soft}^8} {\rm Im}\,{\rm Tr}\,\left[y^{\hat{d}} (y^{\hat{d}})^\dag y^{\hat{d}}\delta m_{\tilde{Q}}^2 \,(\widetilde{A}^{\hat{d}})^\dag \delta m_{\hat{\tilde{\bar{d}}}}^2\right] \\[1ex] &\sim 
	\frac{\alpha_s}{4\pi}\dfrac{v_{d}^3\,v_{u}}{m_{\tilde{g}}m_{\rm soft}^7} {\rm Im}\,{\rm Tr}\,\left[y^{\hat{d}} (y^{\hat{d}})^\dag y^{\hat{d}}\delta m_{\tilde{Q}}^2  \, (y^{\hat{d}})^\dag \delta m_{\hat{\tilde{\bar{d}}}}^2\right] \, ,
	\end{split}
\end{equation}
where $v_{u(d)}\equiv \langle H_{u(d)}\rangle $ are Higgs VEVs and $\widetilde{A}^{\hat{d}}\equiv A^{\hat{d}}+ y^{\hat{d}}\mu \,v_{u}/v_d$.
In the second line, we use $\widetilde{A}^{\hat{d}}\sim y^{\hat{d}}\mu \,v_{u}/v_d\sim y^{\hat{d}}m_{\rm soft}v_{u}/v_d >A^{\hat{d}}$, and the dependence of $m_{\rm soft}$ is estimated using dimensional analysis.
The Feynman diagram of the lowest non-vanishing contribution to $\delta m_{\hat{d}}$ with the CP-violating phase is shown in the right panel of Fig.~\ref{fig:radiative corrections}.
This contribution can be expressed as
\begin{equation}
\begin{split}
     \delta\theta^m_{\hat{d}} &\sim \frac{\alpha_s}{4\pi}\frac{1}{m_{\rm soft}^5}{\rm Im}\,{\rm Tr}\,\left[(y^{\hat{d}} )^{-1} \delta m_{\hat{\tilde{\bar{d}}}}^2\, \widetilde{A}^{\hat{d}} \delta m_{\tilde{Q}}^2 \right] \\[1ex]
     &\sim \frac{1}{16\pi^2}\frac{\alpha_s}{4\pi}\frac{1}{m_{\rm soft}^2}\frac{v_u}{v_d} {\rm Im}\left[  (y^{\hat{d}} )^{-1} \delta m_{\hat{\tilde{\bar{d}}}}^2\,y^{\hat{d}} (y^u)^\dag y^u \right]. \label{eq:quark mass phase}
     \end{split}
\end{equation}
In the second relation, we have used $\widetilde{A}^{\hat{d}}\sim y^{\hat{d}}m_{\rm soft} v_u / v_d$ and $\delta m_{\tilde{Q}}^2\sim (y^u)^\dag y^u m_{\rm soft}^2/(16\pi^2)$, which are dominated
by the radiative correction of the top Yukawa coupling.
Note that $\delta \theta^m_{\hat{d}}$ vanishes for $\delta m_{\hat{\tilde{\bar{d}}}}\sim  y^{\hat{d}} (y^{\hat{d}})^\dag m_{\rm soft}^2/16\pi^2$ because of the hermiticity. Therefore,
the constraint from $\delta m_{\tilde{Q}}^2$ in Eq.~\eqref{eq:flavor nonuniversal soft mass} is subdominant.

In general, when $A$-parameters are not precisely proportional to the corresponding Yukawa coupling,
additional radiative corrections to $\bar{\theta}$ proportional to ${\rm Im}\,{\rm Tr}[(y^{\hat{d}})^\dag A^{\hat{d}}]$
and ${\rm Im}\,{\rm Tr}[(y^{\hat{d}})^{-1}A^{\hat{d}}]$ are generated~\cite{Hiller:2002um}.
However, these contributions vanish for the $A$-parameter given by Eq.~\eqref{eq:flavor nonuniversal soft mass}.
We have also confirmed that radiative corrections to phases of the up-type SM quark masses are not induced by the soft SUSY breaking parameters given by Eq.~\eqref{eq:flavor nonuniversal soft mass}.
Therefore, there is no additional constraint from the up-type sector.

The radiative corrections to the phase of the gluino are subdominant compared to those of the SM quarks
because they are suppressed by powers of $v_{u,d}/m_{\rm soft}$.
Then, the stringent constraint originates from $\delta \theta^m_{\hat{d}}$.
By substituting the expression in Eq.~\eqref{eq:flavor nonuniversal soft mass} into Eq.~\eqref{eq:quark mass phase},
the bound is given by
\begin{align}
(y^{\hat{D}}_{\alpha i} y^{\hat{D}}_{\beta k})^*y^{\hat{D}}_{\alpha k}y^{\hat{D}}_{\beta j}\tan\beta\frac{M_*^2}{M_{\rm CP}^2} \lesssim 10^{-3}.
\label{eq:mediation scale EFT}
\end{align}
In this calculation, we have used $\tan \beta\equiv v_u/v_d \gg1$ and the condition $\delta \theta^m_{\hat{d}}<10^{-10}$.

Finally, let us comment on the case with $M_* > M_{\rm CP}$.
Although the EFT analysis presented above cannot be applied,
we have checked that the similar Feynman diagram gives the most important contribution to $\delta\bar\theta$.
The constraint can be obtained by replacing $M_{\rm CP}^2$ in Eq.~(\ref{eq:mediation scale EFT}) to $M_*^2$.
To summarize, in the case of either $M_* < M_{\rm CP}$ or $M_* > M_{\rm CP}$, $\delta\bar\theta \lesssim 10^{-10}$ is satisfied if
\begin{align}
(y^{\hat{D}}_{\alpha i} y^{\hat{D}}_{\beta k})^*y^{\hat{D}}_{\alpha k}y^{\hat{D}}_{\beta j}\tan\beta \times {\rm min}\left[ \frac{M_*^2}{M_{\rm CP}^2}, 1 \right] \lesssim 10^{-3}.\label{eq:mediation scale}
\end{align}
This constraint will be used in Sec.~\ref{plots}
to determine the parameter space where the strong CP problem, baryon asymmetric Universe,
and DM are explained.

\section{Affleck-Dine baryogenesis via a heavy quark flat direction} \label{sec:ADBG}

Let us consider baryogenesis in the SUSY NB model. 
We will verify that the AD baryogenesis works without introducing further new fields nor CP-violating operators in the model. 

The scalar potential in a supersymmetric model becomes zero if all of the $F$-terms and $D$-terms are zero.
In some cases, there are non-trivial scalar VEV solutions with $F_i=D_a=0$ which can be parameterized by continuous parameters.
The scalar potential is flat in such directions, which are called flat directions.%
\footnote{
In the context of the AD baryogenesis, even if a direction has a Dirac mass from $F$-terms,
it is conventional to call it a flat direction.
}
It has been known that flat directions are characterized by gauge invariant combinations of chiral superfields \cite{Buccella:1982nx, Affleck:1983mk, Affleck:1984xz}.
A part of flat directions in our model are then listed in Table~\ref{flat directions},
where we neglect the Dirac mass term and only write the ones associated with $D$ and/or $\DD$.
The flat directions in the MSSM can be found in Ref.~\cite{Gherghetta:1995dv}. 
The AD baryogenesis is the mechanism that generates $B-L$ asymmetry via the dynamics of a flat direction with a nonzero $B-L$ charge, called an AD field~\cite{Affleck:1984fy, Murayama:1993em, Dine:1995kz}. 
The resulting $B-L$ asymmetry is converted to the baryon asymmetry via the sphaleron effect before the electroweak phase transition~\cite{Kuzmin:1985mm,Fukugita:1986hr}.

\renewcommand{\arraystretch}{1.2}
\begin{table}
\begin{center}
\begin{tabular}{ll}
 &
$B-L$ 
\\
 \hline \hline
$ D \DD $ &$\,\,\,\, 0$ \\ \hline
$ D \dd $ &$\,\,\,\, 0$ \\ \hline
$ \uu \dd \DD $ &$-1$ \\  \hline 
$ Q \DD L $ &$-1$ \\  \hline 
$ \uu \ee D $ &$-1$ \\ \hline
$ QQD $ &$\,\,\,\, 1$ \\ \hline
$ Q\uu Q \DD $ &$\,\,\,\, 0$ \\ \hline
$ \uu \uu \DD \ee $ &$\,\,\,\, 0$ \\ \hline
$ \dd \dd \DD L L $ &$-3$ \\ \hline
\end{tabular}
\end{center}
\caption{\label{flat directions}
Some flat directions associated with $D$ and/or $\DD$ and their $B-L$ charges. 
The Dirac mass term is neglected to define the flat directions. 
}
\end{table}

In this study, we assume that the CP symmetry is not restored by the dynamics of the AD field, to avoid the domain wall problem of spontaneous CP violation. 
In particular, we should not use flat directions including $\dd$ and/or $D$. This is because the CP violating field $\eta$ obtains a large effective mass via the Yukawa term if the flat direction including $\dd$ and/or $D$ obtains a large VEV, and then the CP symmetry is restored in the early Universe. 
For concreteness, 
we identify the $Q \DD L$ flat direction as the AD field to attain the AD mechanism. This flat direction contains a heavy squark and hence provides a unique scenario for the NB model. 
Note that there is no Yukawa term that directly connects $Q$ and $\DD$.
This is in contrast to the $Q \dd L$ flat direction in the MSSM,
which has a renormalizable $F$-term potential from the SM Yukawa interaction.

\subsection{The potential of the AD field}

First, we describe the potential of the AD field during and after inflation. 
We consider the scenario of non-instantaneous reheating after inflation, where inflation is followed by an inflaton-oscillation dominated era (or so-called early matter-dominated era) and then by the radiation dominated era. We denote the Hubble parameter during inflation as $H_{\rm inf}$ and the reheating temperature as $T_{\rm RH}$. 
It is assumed that $H_{\rm inf}$ is larger than the mass of the AD field.

If the AD field has a very flat potential, it may have a large VEV in the early Universe. 
Therefore, higher-dimensional operators 
are important in the stabilization of its VEV at a high-energy scale. 
We may write a non-renormalizable superpotential for the AD field~\cite{Dine:1995kz},
\beq
 W \supset \frac{\lambda}{(N!)^3 \Mpl^{3(N-1)}} \lmk Q \DD L \rmk^{N}, 
\eeq
where the flavor indices have been omitted for notational simplicity. 
The superpotential of the AD field $\phi$ is obtained by identifying $\phi^3 \approx Q \DD L$. 
This superpotential leads to the corresponding $A$-term via the gravity-mediated SUSY breaking effect. 
Although this $A$-term can produce baryon asymmetry,
it is strongly suppressed by a small gravitino mass in our model. 
We thus instead consider the case where  
the above non-renormalizable superpotential is absent owing to an implicit approximate (discrete) R-symmetry. 
For example, we can use the $Z_{4R}$ R-symmetry to forbid the above superpotential for the case of $N=4$. 
In this case, non-renormalizable K\"ahler potentials (rather than superpotentials) are important for the dynamics of the AD field. 
Here, we only consider the minimal terms that are important for our discussion~\cite{Dine:1995kz,Fujii:2002kr,Fujii:2002aj}: 
\beq
 && K \supset K_H + K_{A} \, ,
 \\[1ex]
 && K_H = -\frac{c_H}{\Mpl^2} I^\dagger I \phi^\dagger \phi
 + \frac{c_K}{\Mpl^6} I^\dagger I \lmk Q^\dagger \DD^\dagger L^\dagger \rmk \lmk Q \DD L \rmk, \\[1ex]
 && K_A =  
 -\frac{c_A}{(N!)^3 \Mpl^{3N}} I^\dagger I \lmk Q \DD L \rmk^{N} 
 + {\rm h.c.}
\eeq
where $c_{H,K,A}$ are $\mathcal{O}(1)$ real constants. 
In the first term, $\phi^\dagger \phi$ can be understood as a collective notation for $Q^\dagger Q$, $\DD^\dagger \DD$, and $L^\dagger L$. 
The field whose $F$-term dominates the energy density of the Universe during inflation and inflaton-oscillation dominated era is collectively denoted as $I$, such that $\abs{F_I}^2 \sim 3 H^2(t) \Mpl^2$, where $H(t)$ is the Hubble parameter.%
\footnote{
Here, we have assumed that inflation is driven by an $F$-term potential for concreteness. 
Even if inflation is driven by a $D$-term potential~\cite{Binetruy:1996xj,Halyo:1996pp},
our result is not significantly changed~\cite{Kolda:1998kc,Enqvist:1998pf,Enqvist:1999hv,Kawasaki:2001in}. 
}
We include 
the factorial prefactors in the denominator of the non-renormalizable K\"ahler potential $K_A$
such that the cutoff scale is of the order of $\Mpl$ by canceling the combinatorial factor.
Only the smallest dimensional $U(1)_{B-L}$-breaking term, such as $K_A$, is important.
Note that the K\"ahler potential violates $B-L$ symmetry but maintains CP symmetry. Thus, it is consistent with the NB mechanism.

With the non-renormalizable K\"ahler potentials,
the potential of the AD field $\phi^3 \approx Q \DD L$ during inflation and the inflaton-oscillation dominated era is given by: 
\beq
 V(\phi) = m_\phi^2 \abs{\phi}^2 - c_H H^2 (t) \abs{\phi}^2  
 + \frac{c_K H^2(t)}{\Mpl^{4}} \abs{\phi}^{6} 
 - 
 \lmk \frac{c_A H^2(t)}{(N!)^3 \Mpl^{3N-2}} \phi^{3N} + {\rm c.c.} \rmk. 
 \label{V}
\eeq
Here, some $\mathcal{O}(1)$ factors are absorbed into $c_{H,K,A}$.%
\footnote{
A factor of $3$ for $H^2 \abs{\phi}^2$ originates from the supergravity potential, even if $K_H$ is absent. It is absorbed into $c_H$ in \eq{V} for notational simplicity. 
}
To realize the AD mechanism, we assume $c_H > 0$ and $c_K >0$. 
Hereafter, we assume $c_A >0$ for simplicity 
although a negative $c_A$ is also allowed in our scenario.%
\footnote{
If the VEV of $\phi$ is as large as the Planck scale, as we will verify shortly, the thermal effect on the AD field is negligible~\cite{Dine:1995kz,Allahverdi:2000zd,Anisimov:2000wx,Fujii:2001zr}. 
}
Note again that there is no CP violating term in the potential. 
In the first term, $m_\phi$ is the mass of the AD field 
that arises after turning on the Yukawa interaction and the Dirac mass term for the heavy quark. 
Since $D$ and $\DD$ have the Dirac mass term, 
the mass of the AD field is given by $m_\phi \sim M_D$. 
This is a bare mass term in the superpotential so that it is not suppressed even for an energy scale larger than the messenger mass. This is in contrast to other MSSM flat directions, for which the soft mass terms are suppressed at an energy scale above the messenger scale~\cite{deGouvea:1997afu}.

\subsection{The dynamics}

Let us now discuss the dynamics of the AD field. 
For this purpose, we decompose the AD field into radial and phase directions, such as $\phi = \cphi e^{i \theta} / \sqrt{2}$. 
The equation of motion of the AD field is then given by 
\beq
 \ddot{\cphi} + 3 H \dot{\cphi} - \dot{\theta}^2 \cphi + \frac{\del V}{\del \cphi} = 0 \, , 
 \label{eom1}
 \\[1ex]
 \ddot{\theta} + 3 H \dot{\theta} + 2 \frac{\dot{\cphi}}{\cphi} \dot{\theta}
 + \frac{1}{\cphi^2} \frac{\del V}{\del \theta} = 0 \, . 
 \label{eom2}
\eeq
We consider the case where the Hubble parameter during inflation $H(t) \simeq H_{\rm inf}$ is larger than $M_D$. Specifically, we assume that: 
\beq
 \sqrt{c_H} H_{\rm inf} \gtrsim m_\phi \sim M_D \, . 
 \label{eq:H_inf_bound1}
\eeq
Then, we can neglect $m_\phi$ in \eq{V} and the AD field remains at a large VEV during inflation: 
\beq
 \la \cphi \ra \simeq  \lmk \frac{4c_H }{ 3 c_K} \rmk^{1/4} \Mpl \, . 
 \label{phiosc}
\eeq
The condition~(\ref{eq:H_inf_bound1}) is necessary to obtain the large AD field VEV in Eq.~(\ref{phiosc}).
One might worry about the domain wall problem from the restoration of CP symmetry because of a large $H_{\rm inf}$, however, this problem can be avoided by introducing a negative Hubble induced mass for the $\eta$ field.
Note that the effect of the last term in Eq.~(\ref{V}) on $\langle\varphi\rangle$ is negligible because of the suppression by $1/(N!)^3$.
Regarding the phase direction, 
the solution to \eq{eom2} is given by 
\beq
 \theta \simeq \theta_{\rm min} 
+ \lmk \theta_{\rm ini} - \theta_{\rm min} \rmk
 \exp\lkk - \lmk  \frac{r_\theta}{3} \rmk H_{\rm inf} t \rkk, 
\eeq
for $r_\theta \ll 1$ 
and $|\theta_{\rm ini} - \theta_{\rm min}| \lesssim 1/3N$, 
where $\theta_{\rm ini}$ is the initial phase and $\theta_{\rm min}$ ($= (2\pi/3N)i$, $i=0,1,2,\dots, 3N-1$) represents the potential minima of the phase direction. 
We have also defined
\beq
 r_\theta &\equiv& \left. \frac{\del^2 V}{\cphi^2 \del \theta^2} \frac{1}{H_{\rm inf}^2} \right\vert_{\phi = \la \phi \ra}
 \\
 &\sim& \lmk \frac{(3 N)^2 c_A}{(N!)^3} \rmk \lmk \frac{c_H }{ 3 c_K} \rmk^{(3N-2)/4}. 
\eeq
We consider the case in which the phase direction $\theta$ does not reach the potential minimum during inflation. 
Here, the inflation should last longer than $t \sim H_{\rm inf}^{-1} N_e$,
where $N_e$ ($\sim 50$) is the e-folding number of inflation. 
This implies that for the phase direction not to reach the potential minimum during inflation,
$r_\theta$ should be 
much smaller than $3/N_e$ ($\lesssim 1/20$). 
It can be verified that $r_\theta$ is much smaller than $\mathcal{O}(0.01)$ for $N \ge 3$, depending on $c_{H,K,A}$ ($=\mathcal{O}(1)$). 
For example, $r_\theta \simeq 3 \times 10^{-3}$ for $N=3$ with $5c_H/c_K = c_A = 1$ 
whereas $r_\theta \simeq 7 \times 10^{-4}$ for $N=4$ with $c_H = c_A = c_K =1$. 
Hence, the phase of the AD field is almost massless during inflation and stays at a certain phase for a sufficiently long e-folding number. We denote the phase of the AD field at the end of inflation as $\theta_{\rm ini}$ ($\in (-\pi, \pi)$). 
In general, $\theta_{\rm ini}$ is nonzero and is expected to be $\mathcal{O}(1)$ without fine-tuning. 
Therefore, CP symmetry is spontaneously broken by the AD field during inflation. 
This is a source of CP violation that is required to generate the $B-L$ asymmetry.

After inflation, the Hubble parameter decreases and eventually becomes comparable to $m_\phi/\sqrt{c_H}$. Then, the AD field starts to oscillate around $\cphi = 0$. 
We denote the Hubble parameter and the time at the onset of the oscillation as $H_{\rm osc}$ ($\simeq m_\phi / \sqrt{c_H}$) 
and $t_{\rm osc}$ ($= 2/3H_{\rm osc}$), respectively. 
The amplitude of the oscillation $\cphi_{\rm osc}$ is again given by \eq{phiosc}. 
Simultaneously, the phase direction is kicked by the last term in \eq{V},
and the AD field starts to rotate in the phase space. 
Let us see how $B-L$ number asymmetry is generated from the time evolution of the AD field.
The Noether current of $B-L$ charge is $j_{B-L,\mu} = q_{B-L}(i \phi \partial_\mu \phi^\dagger - i \phi^\dagger \partial_\mu \phi) = q_{B-L} \varphi^2 \partial_\mu\theta$, where $q_{B-L}$ ($=-1$) is the $B-L$ charge of the AD field. We obtain $B-L$ charge density as
\beq
 n_{B-L} = q_{B-L} \cphi^2 \dot{\theta} \, . \label{eq:nBL definition}
\eeq
We can see, provided that the AD field carries the nonzero $B-L$ charge, its rotation generates $B-L$ asymmetry.
By using the equations of motion (\ref{eom1}), (\ref{eom2}), we obtain the time evolution of $n_{B-L}$ defined in Eq.~(\ref{eq:nBL definition}) as
\beq
 \frac{1}{a(t)^3} \frac{\del}{\del t} 
 \lmk a(t)^3 n_{B-L} \rmk 
 = - q_{B-L} \frac{\del V}{\del \theta} \, .
 \label{eq:nbl}
\eeq
We can see that the explicit $U(1)_{B-L}$ breaking term $\partial V/\partial \theta$ works as a source term for $B-L$ asymmetry.

\begin{figure}[t]
\begin{center}
\includegraphics[clip, width=11cm]{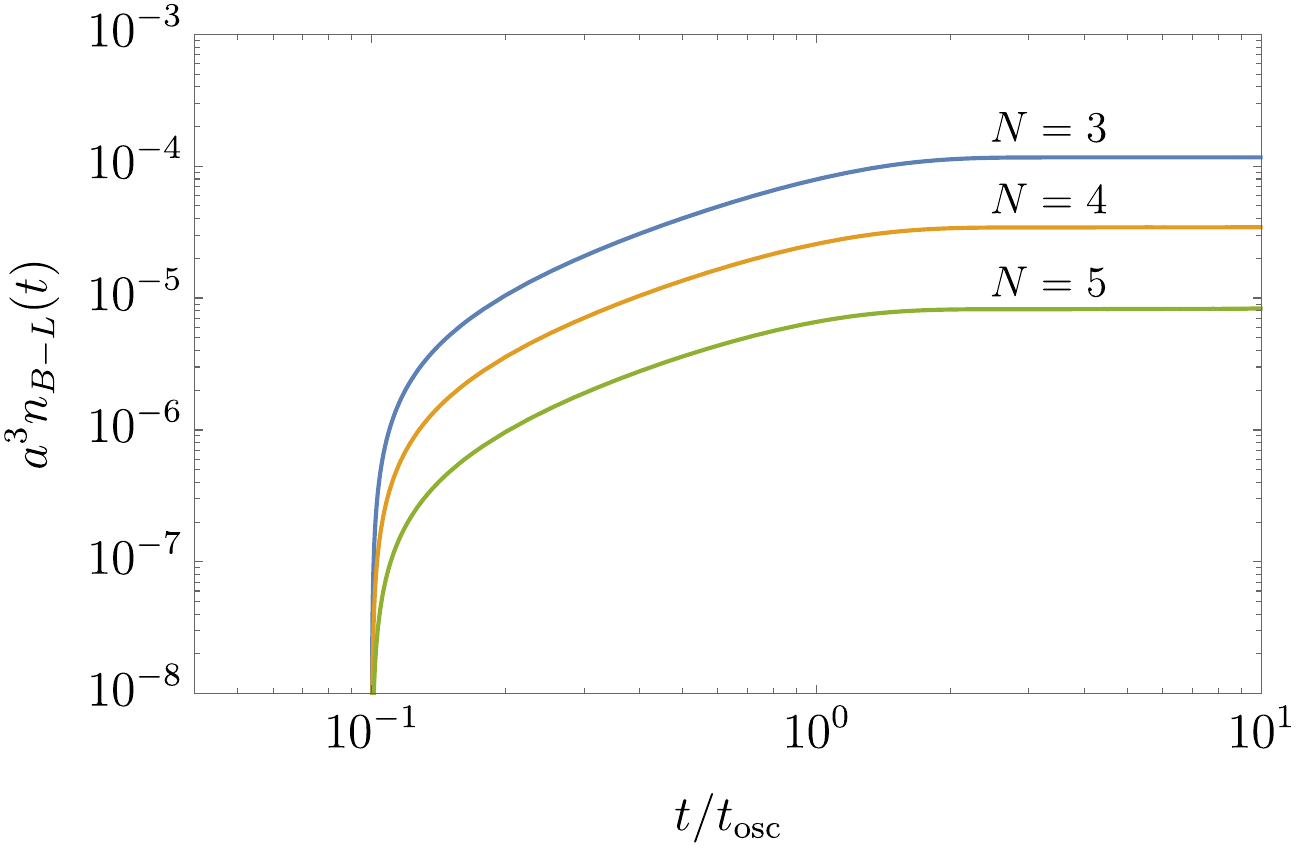}
\end{center}
\caption{ Time evolution of $B-L$ charge density of the AD field in a comoving volume. 
We choose $N=3$ with $c_K = 5$ (blue curve), $N=4$ with $c_K = 1$ (yellow curve), and $N=5$ with $c_K = 1/3$ (green curve).
We take $3 N \theta_{\rm ini} = c_A = c_H = - q_{B-L} = 1$ for each case. 
\label{fig:nBL}
}
\end{figure}

The numerical results of time evolution of $B-L$ charge density of the AD field in a comoving volume
are shown in Fig.~\ref{fig:nBL}, 
where we choose $N=3$ with $c_K = 5$, $N=4$ with $c_K = 1$, and $N=5$ with $c_K = 1/3$. 
We take $3 N \theta_{\rm ini} = 1$ for each $N$. 
Other $\mathcal{O}(1)$ parameters are set to unity. 
It can be seen that the $B-L$ asymmetry is generated around $t = t_{\rm osc}$ and its comoving density is conserved for $t \gg t_{\rm osc}$. 
We can also obtain the analytic formula by approximating the time integral in Eq.~\eqref{eq:nbl}
as $\int dt \cphi^{3N} \sin (3 N \theta) \sim t_{\rm osc} \cphi_{\rm osc}^{3N} \sin (3 N \theta_{\rm ini})$,
\beq
 n_{B-L}(t) = q_{B-L} \epsilon  \, m_\phi \cphi_{\rm osc}^2 \frac{a (t_{\rm osc})^3}{a(t)^3} \, . 
 \label{nb}
\eeq
Here, the ellipticity $\epsilon$ is given by 
\beq
 \epsilon =
- c_\epsilon \sin \lmk 3 N \theta_{\rm ini} \rmk
\lmk \frac{4N c_A}{(N!)^3 \sqrt{c_H}} \rmk \lmk \frac{c_H }{ 3 c_K} \rmk^{(3N-2)/4}  ,
 \label{epsilon}
\eeq
where $H_{\rm osc} = 2/(3 t_{\rm osc}) = m_\phi / \sqrt{c_H}$, and $c_\epsilon$ is the $\mathcal{O}(1)$ numerical factor. 
This is approximately proportional to the initial phase and ratio in the potential curvature between the radial and phase directions. 
Using our numerical results, we can verify that Eqs.~(\ref{nb}) and (\ref{epsilon}) with $c_\epsilon = 0.5\,\text{-}\,1$ gives the correct value for the parameter in the region of interest (namely, for $N, \theta_{\rm ini},
c_{H,K,A} = \mathcal{O}(1)$).

The amplitude of $\phi$ decreases owing to the cosmic expansion, and the non-renormalizable $B-L$ breaking operators become ineffective after the onset of the oscillation. 
It can be observed from \eq{eq:nbl} that 
the $B-L$ number-to-entropy ratio is conserved for $t \gg t_{\rm osc}$ and is given by 
\begin{eqnarray}
\frac{n_{B-L}}{s} 
&=& \left. \frac{n_{B-L}}{4\rho/3T} \right\vert_{T= T_{\rm RH}} 
= \left. \frac{3T_{\rm RH}}{4} \frac{n_{B-L}}{3 H^2 \Mpl^2} \right\vert_{t = t_{\rm osc}}
\nonumber \\[1ex]
&=& q_{B-L} \epsilon c_H \frac{T_{\rm RH}}{4m_\phi} \frac{ \cphi_{\rm osc}^2}{ \Mpl^2}
\nonumber \\[1ex]
&=&
- \sin \lmk 3 N \theta_{\rm ini} \rmk q_{B-L} \frac{T_{\rm RH}}{m_\phi}
\lmk \frac{ 2N c_A c_\epsilon \sqrt{c_H}}{ (N!)^3 } \rmk  \lmk \frac{ c_H }{ 3 c_K} \rmk^{3N/4} .
\label{B-L}
\end{eqnarray}
We have used Eqs.~(\ref{phiosc}), \eqref{nb}, (\ref{epsilon}). 
Here, $T_{\rm RH}$ is the reheating temperature, that is, the temperature at the beginning of the radiation-dominated era. 
We do not assume instantaneous reheating, but 
the case in which 
the inflaton-oscillation dominated era 
is followed by 
the radiation-dominated era is considered. 
For example, the $B-L$ number-to-entropy ratio is given by 
\beq
 \frac{n_{B-L}}{s} 
&\simeq& 3 \times 10^{-10} \, \sin \lmk 3 N \theta_{\rm ini} \rmk 
\lmk \frac{T_{\rm RH}}{10^3 \GeV} \rmk \lmk \frac{m_\phi}{10^{8} \GeV} \rmk^{-1}, 
\label{YbN3}
\eeq
for $N = 3$ with $c_A = c_H = c_K/5 = 1$ (in which case, $c_\epsilon \simeq 1.1$), 
and 
\beq
 \frac{n_{B-L}}{s} 
&\simeq& 2 \times 10^{-10} \, \sin \lmk 3 N \theta_{\rm ini} \rmk 
\lmk \frac{T_{\rm RH}}{10^3 \GeV} \rmk \lmk \frac{m_\phi}{10^{8} \GeV} \rmk^{-1}. 
\label{YbN4}
\eeq
for $N = 4$ with $c_A = c_H = c_K = 1$ (in which case $c_\epsilon \simeq 0.97$). 
Note that $m_\phi \sim M_D$, which is much larger than the visible sector superpartner mass scale.

\subsection{Dissipation into thermal plasma}

After the $B-L$ asymmetry is generated by the dynamics of the AD field, it should be dissipated into thermal plasma. 
If the dissipation is completed before the electroweak phase transition, the $B-L$ asymmetry is converted into a combination of baryon and lepton asymmetries via the electroweak sphaleron. 
We will verify that this is indeed the case in our model. 

Since the amplitude of $\phi$ is as large as the Planck scale, 
the energy density of its oscillation is comparable to that of the inflaton oscillation. 
If the reheating is completed before the dissipation of $\phi$, its energy density dominates the Universe. 
In this case, $T_{\rm RH}$ in \eq{B-L} must be replaced by the decay temperature of the AD field. 
Since the decay rate of the AD field is quite large, its decay temperature cannot be as low as the scale of interest. 
Thus, it is necessary to check whether
the dissipation temperature is higher than the reheating temperature.

If the AD field has a very large field value, it cannot decay into particles that directly couple to the AD field at the tree level. Instead, the AD field can dissipate into light particles via loop diagrams. The rate of dissipation into gauge fields via loop diagrams is estimated as (see, for example, Refs.~\cite{Mukaida:2012qn,Mukaida:2012bz})
\beq
\Gamma_{\rm diss}(t) \sim 10^{-2} \alpha^2 y^2 \frac{m_\phi^3}{(y \phi)^2} \, , 
\eeq
where $\alpha$ and $y$ generically represent the fine-structure constants for SM gauge interactions and Yukawa interactions for the AD field, respectively. 
The dependence of the Yukawa coupling is cancelled between the denominator and numerator. 
Here, we implicitly assume $m_\phi > \alpha T$ and $y \phi \gg m_\phi$, 
which is satisfied during the regime of interest. We compare the dissipation rate with the Hubble expansion rate and obtain 
\beq
\frac{\Gamma_{\rm diss}(t)}{H(t)} \sim 
10^{-2} \alpha^2 \lmk \frac{m_\phi}{\Mpl} \rmk^2 
\lmk \frac{\Mpl}{\phi_{\rm osc}} \rmk^2 \lmk \frac{m_\phi}{H(t)} \rmk^3, 
\eeq
where we have considered that 
the amplitude of $\phi$ decreases as $\propto H(t)$. 
The ratio becomes larger than $\mathcal{O}(1)$ 
by the completion of the reheating 
for a typical parameter space of interest. 
This means that the AD field dissipates into the thermal plasma well before the completion of the reheating, at which $H(T_{\rm RH}) \sim 10^{-12} \GeV$ for $T_{\rm RH} \sim 1\TeV$. Therefore, we conclude that the AD field does not dominate the Universe, and thus, the calculation of the baryon asymmetry can be justified.

In the preceding discussions, we have assumed that the AD field does not form Q-balls~\cite{Coleman:1985ki},
which are localized condensation with a large number of $B-L$ charges. 
If the potential of the AD field is less than quadratic, spacial perturbations of the AD field increase and eventually Q-balls form after the onset of the oscillation~\cite{Kusenko:1997zq, Kusenko:1997si, Enqvist:1997si, Enqvist:1998en,Kasuya:1999wu,Kasuya:2000wx,Kasuya:2001hg}. 
In our case, 
the mass of the AD field $m_\phi$ has logarithmic quantum corrections via 
the renormalization group running 
and its potential can be written as 
\beq
 V(\phi) \simeq m_\phi^2 \lkk 1 + K \log \lmk \frac{\abs{\phi}}{\Mpl} \rmk \rkk \abs{\phi}^2, 
\eeq
where $K$ is a small parameter determined by the quantum corrections. 
Here, we neglect the higher-dimensional operators because they are not important for $\abs{\phi} \ll \Mpl$. 
If $K$ is negative, the potential is shallower than the quadratic and Q-ball forms after the onset of the oscillation. 
The sign of $K$ depends on the value of the quantum corrections from the Yukawa interactions compared to those from the gauge interactions. If the Yukawa coupling is sufficiently small, then $K$ is negative. Depending on the value of the Yukawa coupling, the Q-balls may or may not form in our scenario.

Q-balls are localized condensations with a high number density, and
their decay rate into fermions is strongly suppressed because the flux of fermions is saturated by the Pauli exclusion principle on the surface of the Q-balls~\cite{Cohen:1986ct}. 
In the standard AD baryogenesis, 
Q-balls are therefore long-lived (or stable), and the thermal history and the calculation of baryon asymmetry may drastically change. In our case, however, 
the AD field consists of the heavy squark, which can decay into 
light bosons (the MSSM squarks and Higgs), without being suppressed by the Pauli exclusion principle. Rather, their decay rate is enhanced exponentially by the Bose enhancement effect~\cite{Hertzberg:2010yz,Kawasaki:2013awa} similarly to the preheating process~\cite{Kofman:1994rk,Kofman:1997yn,Shtanov:1994ce,Greene:1997fu,Traschen:1990sw}. 
This means that even if Q-balls are formed, they quickly dissipate into light particles. Therefore, we conclude that the formation of the Q-balls does not affect the calculations in our case.

The AD field dissipates before the electroweak phase transition, 
and the $B-L$ asymmetry is converted to a combination of baryon and lepton asymmetries via the electroweak sphaleron. 
This is the same even if the reheating temperature is lower than the electroweak scale because the electroweak sphaleron is efficient within the ambient plasma even before the reheating is completed (see, for example, Ref.~\cite{Mukaida:2015ria}). 
In any case, the resulting baryon-to-entropy ratio is given by~\cite{Harvey:1990qw}
\beq
 Y_B = \frac{8}{23} \frac{n_{B-L}}{s} \, . 
\eeq
The observed value is $9 \times 10^{-11}$~\cite{Planck:2018vyg}. 
Thus, using \eq{B-L}, we can realize the baryon asymmetric Universe in the SUSY NB model.

\subsection{Quantum fluctuations}

The phase of the AD field is almost massless, so that it acquires quantum fluctuations during inflation,
\beq
 \abs{\delta \theta_{\rm ini}} \simeq \frac{H_{\rm inf}}{2 \pi \la \cphi \ra} \, . 
 \label{deltatheta}
\eeq
The produced baryon asymmetry depends on $\theta_{\rm ini}$ (see \eq{B-L}), 
and hence the fluctuations 
result in isocurvature fluctuations of baryon asymmetry~\cite{Enqvist:1998pf,Enqvist:1999hv,Kawasaki:2001in,Kasuya:2008xp,Harigaya:2014tla}:  
\beq
 S_{b \gamma} \equiv 
 \frac{\delta Y_B}{Y_B} \simeq 3N \cot ( 3N \theta_{\rm ini} ) \delta \theta_{\rm ini} \, .
 \label{Sbgamma}
\eeq
Since the observed density perturbations are predominantly adiabatic, 
the isocurvature fluctuations are constrained by the Planck results, such as 
$S_{b \gamma} \lesssim 
5.0 \times 10^{-5}$. 
Together with Eqs.~(\ref{phiosc}) and (\ref{deltatheta}), we obtain 
\beq
 H_{\rm inf} \lesssim 8.1 \times 10^{14} \GeV \lmk \frac{\tan (3 N \theta_{\rm ini})}{3N} \rmk \lmk \frac{c_H}{c_K} \rmk^{1/4}. 
\eeq
This bound will be improved by a factor of a few by future observations of the cosmic microwave background (CMB), such as LiteBIRD~\cite{Matsumura:2013aja,Matsumura:2016sri} and CORE~\cite{CORE:2016ymi}.

The energy scale of inflation is constrained by the measurement of the tensor modes for CMB temperature anisotropies. 
The upper bound on the tensor-to-scalar ratio is approximately given by $r \simeq 0.036$~\cite{BICEP:2021xfz},
which implies that 
\beq
 H_{\rm inf} \simeq 4.6 \times 10^{13} \GeV \lmk \frac{r}{0.036} \rmk^{1/2}. 
 \label{tensortoscalar}
\eeq
The future CMB missions, such as  LiteBIRD~\cite{Matsumura:2013aja,Matsumura:2016sri},  CORE~\cite{CORE:2016ymi}, and CMB-S4~\cite{CMB-S4:2020lpa}, will measure $r$ with uncertainty of $\mathcal{O}(10^{-3})$. 
Combining \eq{Sbgamma} with \eq{tensortoscalar}, we predict a consistency relation between the isocurvature fluctuations and the tensor-to-scalar ratio such as 
\beq
 \lmk \frac{S_{b \gamma}}{5 \times 10^{-5}} \rmk = \lmk \frac{\theta_{\rm ini}}{0.06} \rmk^{-1} \lmk \frac{\tan (3 N \theta_{\rm ini})}{3N \theta_{\rm ini}} \rmk^{-1} \lmk \frac{c_H}{c_K} \rmk^{-1/4}
 \lmk \frac{r}{0.036} \rmk^{1/2}.
\eeq
If $H_{\rm inf}$ is close to the upper bounds, 
both isocurvature fluctuations and the tensor-to-scalar ratio would be observed in future.

\section{Summary plots and other constraints}\label{plots}

We now combine the results and constraints obtained from the previous sections to determine the parameter space in which the strong CP problem, baryon asymmetric Universe, and DM are explained. 
However, it is still needed to take account of some constraints in cosmological scenarios, such as the domain-wall formation and the gravitino overproduction problem. 
We first explain these issues and then show summary plots. 

\subsection{The domain-wall problem}

Let us consider the constraint on the maximum temperature of the Universe after inflation.
If CP symmetry is spontaneously broken after inflation, domain walls are formed.
The domain walls soon dominate the Universe and lead to a highly inhomogeneous Universe. 
To avoid this problem, we consider a scenario in which CP symmetry is spontaneously broken before inflation and is never restored after inflation. 
For this, the maximal temperature after inflation, $T_{\rm max}$, must be lower than the mass of $\eta$,
which is expected to be at the order of the VEV of $\eta$. 
The maximal temperature is estimated as 
\beq
 \label{eq:T_max}
 T_{\rm max} \sim \lmk \frac{\Gamma_I}{H_{\rm inf}} \rho_{\rm inf} \rmk^{1/4} \sim \lmk T_{\rm RH}^2 H_{\rm inf} \Mpl \rmk^{1/4}, 
\eeq
where $\rho_{\rm inf}$ ($\simeq 3 H_{\rm inf}^2 \Mpl^2$) and $\Gamma_I$ ($\sim T_{\rm RH}^2 / \Mpl$) are the energy density during inflation and the inflaton decay rate, respectively. 
We require 
\beq
\label{eq:T_max_bound}
 T_{\rm max} \lesssim \la \eta_\alpha \ra \sim \frac{M_{\rm CP} }{ y^D} \, .
\eeq
Note that \eq{eq:T_max} overestimates the maximal temperature because it usually requires a finite time for inflaton-decay products to be thermalized~\cite{Mukaida:2015ria}. Therefore, the preceding condition should be regarded as a conservative upper bound on the reheating temperature.

\subsection{The gravitino problem and Lyman-$\alpha$ constraint}

Gravitino overproduction results in a stringent constraint on the reheating temperature.
As we have discussed in Sec.~\ref{sec:model}, the gravitino mass must be smaller than $\mathcal{O}(10\,\text{-}\,100) \KeV$ to solve the strong CP problem in the NB model. 
In this case, 
if the temperature of thermal plasma is higher than the mass scale of the MSSM SUSY particles $m_{\rm soft}$, 
the stable gravitino is in thermal equilibrium via the decay and inverse decay of the MSSM SUSY particles. 
If the gravitino mass $m_{3/2}$ is larger than $\mathcal{O}(1)\KeV$, and if $T_{\rm RH} \gg m_{\rm soft}$, the gravitino abundance is larger than the DM abundance. This is known as the gravitino problem.
One may consider a lighter gravitino, which is favored for the NB model, as explained in the previous section. 
In this case, the reheating temperature can be arbitrarily large. 
However, the free-streaming constraint by the Lyman-$\alpha$ forest data excludes
the possibility that such a light gravitino is DM.

Instead, one can consider a scenario in which the reheating temperature is lower than the mass of the lightest MSSM SUSY particle. 
In this case, the gravitino is diluted by the reheating process even though it is produced during the inflaton-oscillation dominated era. 
Specifically, 
the gravitino abundance is given by 
\begin{align}
 \frac{\rho_{3/2}}{s} &\simeq 
 \frac{3 T_{\rm RH} }{4} \lmk \frac{\rho_{3/2}}{\rho_{\rm rad}} \frac{\rho_{\rm rad}}{\rho_{\rm tot}}  \rmk_{T = T_{\rm dec,3/2}}
 \nonumber \\[1ex]
 &\simeq m_{3/2} \frac{45 \zeta(3) }{2 \pi^4 g_*}  
 \lmk \frac{T_{\rm RH}}{T_{\rm dec,3/2}} \rmk^5, 
\end{align}
where $\rho_{\rm rad} / \rho_{\rm tot} = (2/3) (T_{\rm RH} / T_{\rm dec,3/2})^4$ in the second line. 
Here, $T_{\rm dec,3/2}$ ($\sim m_{\rm soft}$) denotes the decoupling temperature of the gravitino, and
$g_*$ is the number of the effective relativistic degrees of freedom at $T = T_{\rm dec,3/2}$. 
We can choose $T_{\rm RH}$ to explain the DM abundance by the gravitino, such as 
\beq
 T_{\rm RH} \simeq 0.27 \, T_{\rm dec,3/2}
 \lmk \frac{m_{3/2}}{100 \KeV} \rmk^{1/5} ,
\label{eq:gravitino_DM}
\eeq
where we have used $\rho_{3/2} /s \simeq 0.4 \eV$ for the observed DM density 
and $g_* \simeq 106.75$. 
The reheating temperature must be lower than $m_{\rm soft}$, but it should not be too small because the suppression factor
in the gravitino abundance has a power of five. 
In addition, the free-streaming velocity of the gravitino must be suppressed. 
The Lyman-$\alpha$ constraint reads~\cite{Viel:2005qj,Irsic:2017ixq} (see also Ref.~\cite{Dekker:2021scf}) 
\beq
 m_{3/2} \gtrsim 5.3 \KeV. 
\label{eq:Lyman_alpha}
\eeq
Our results (\ref{YbN3}) and (\ref{YbN4}) show that a sufficient baryon asymmetry can be successfully produced, while satisfying the constraints on $T_{\rm RH}$ and $m_{3/2}$. 
We have found that the Dirac mass of the heavy quark should be $10^{9} \GeV$ or smaller, depending on the $T_{\rm RH}$ and $\mathcal{O}(1)$ constants. This is consistent with the constraints discussed in Sec.~\ref{radiative} to suppress quantum corrections to the strong CP phase.

\subsection{Summary plots}

We now summarize our results and show some plots. 
The strong CP phase is given by the contributions from
the Planck-suppressed operators in \eq{eq:CP-violating mu-term}, 
anomaly-mediated effect in \eq{eq:anomaly mediation}, 
and radiative corrections in \eq{eq:mediation scale}. 
They are summarized as
\beq
 &&\bar{\theta} = \bar{\theta}_{\rm Planck} 
 + \bar{\theta}_{\rm anomaly} + \bar{\theta}_{\rm radiative} \, ,
 \\[1.5ex]
 &&\bar{\theta}_{\rm Planck} \sim 10^{-2} \left( \frac{{\rm Re}\, (\mu) \tan\beta}{M_{\rm Pl}}\right)^{-1} 
 \lmk \frac{\langle \eta_\alpha \rangle }{\Mpl} \rmk^N ,
 \\[1ex]
 &&\bar{\theta}_{\rm anomaly} \sim \frac{\alpha_s}{4\pi} \frac{m_{3/2}}{m_{\tilde{g}}} \, ,
 \label{eq:theta_anomaly}
 \\[1ex] 
 &&\bar{\theta}_{\rm radiative} \sim 
 10^{-7} (y^{\hat{D}}_{\alpha i} y^{\hat{D}}_{\beta k})^*y^{\hat{D}}_{\alpha k}y^{\hat{D}}_{\beta j} 
 \tan \beta
 \times {\rm min}\left[ \frac{M_*^2}{M_{\rm CP}^2},1 \right]. 
 \label{eq:theta_radiative}
\eeq
To show the results, 
we just sum up these contributions with equal signs without fine-tunings. The overall sign can be either positive or negative. 
To explain the observed baryon asymmetry, we 
require that $Y_{B-L}$ is given by \eq{YbN3} (for $N=3$) or \eq{YbN4} (for $N=4$) and is approximately $3 \times 10^{-10}$. 
The Hubble parameter during inflation should satisfy Eqs.~(\ref{eq:H_inf_bound1}) and (\ref{eq:T_max_bound}). 
We can explain the observed DM abundance by the thermally produced gravitino as presented in \eq{eq:gravitino_DM},
for which case \eq{eq:Lyman_alpha} must be satisfied to avoid the Lyman-$\alpha$ constraint.

\begin{figure}[t]
\begin{center}
\includegraphics[clip, width=11cm]{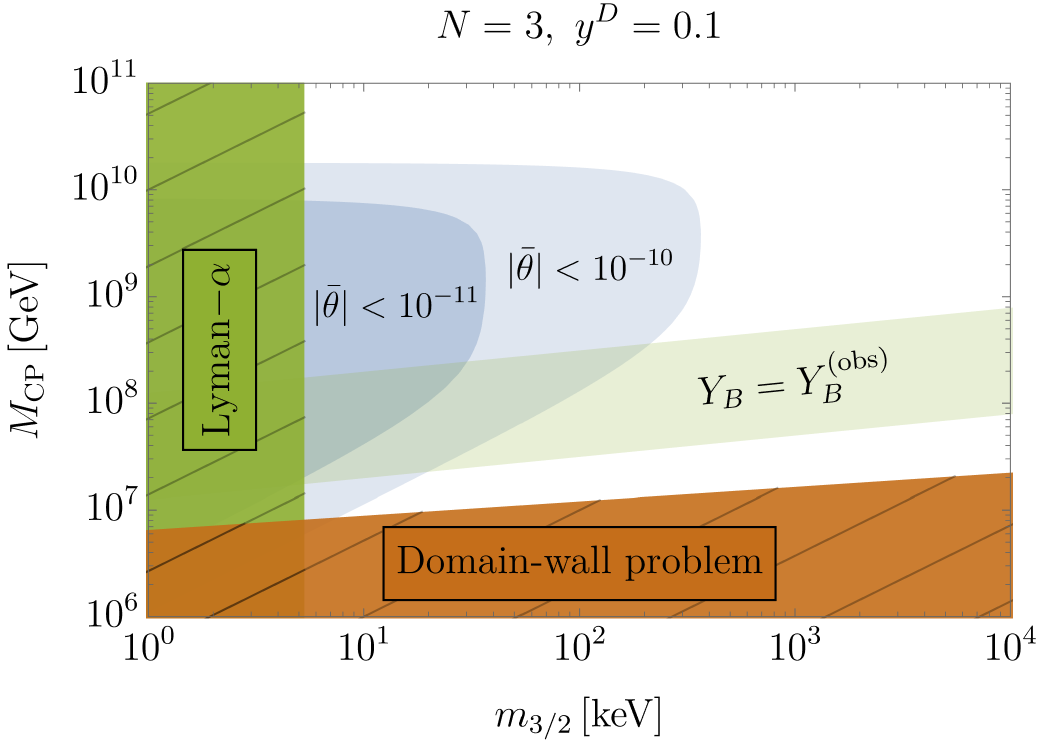}
\\
\vspace{1cm}
\includegraphics[clip, width=11cm]{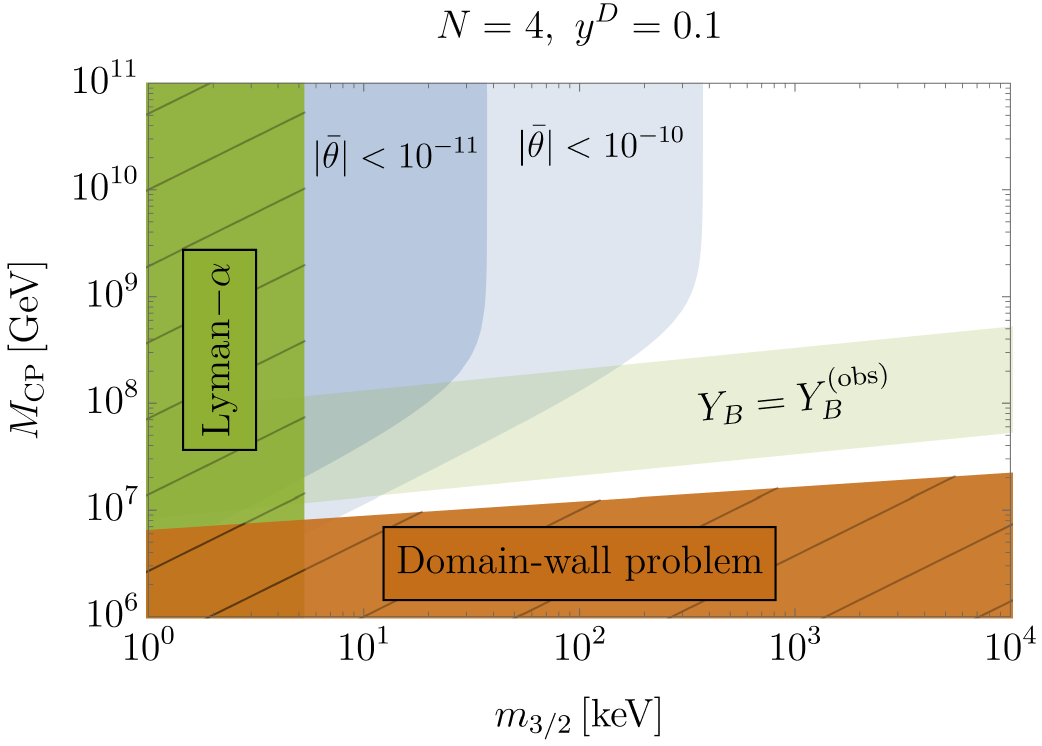}
\end{center}
\caption{The prediction of $\bar{\theta}$ on $(m_{3/2},M_{\rm CP})$-plane for $m_{\rm soft}=10\,{\rm TeV},~y^D=0.1$ with $N=3$ (upper panel) and $N=4$ (lower panel). 
We take the reheating temperature such that the gravitino abundance is equal to the observed DM abundance. 
The meshed green- and brown-colored regions are excluded because of the Lyman-$\alpha$ constraint and by the domain wall formation, respectively. 
The light blue shaded regions show 
$\bar{\theta}<10^{-10},10^{-11}$, whereas
the light green shaded region shows $Y_B = Y^{({\rm obs})}$. 
\label{fig:summary1}
}
\end{figure}

Let us summarize the parameters in our model. 
As explained in Sec.~\ref{sec:model}, $M^2_D \sim B^\dag B$ should be satisfied to obtain the observed CKM matrix.
Then, we take $M_{\rm CP}= M_D= y^D\langle \eta_\alpha \rangle$ 
and $y^{D}\sim y^{\hat{D}}\sim y^{\hat{D}\hat{\bar{D}}}$ for simplicity. 
The messenger scale $M_*$, which enters in \eq{eq:mediation scale}, is given by $M_* = (\alpha/4\pi) \la F \ra / m_{\rm soft}$, where $\la F \ra = \sqrt{3} \Mpl m_{3/2}$ is the SUSY breaking scale and $\alpha$ represents
the fine structure constants of the SM gauge couplings. 
The messenger scale and the SUSY breaking scale must satisfy the relation $\la F \ra < M_*^2$
for the scalar component of a messenger chiral superfield to stay at the origin of the potential. 
We take $\alpha = 1/30$, $\mu = m_{\rm soft}$, and $\tan \beta =10$ as a benchmark point,
although the dependence on $\tan \beta$ is very weak. 
In summary, we consider $m_{\rm soft},m_{3/2},M_{\rm CP},y^D$, and $N$ as free parameters in the model. 
Our prediction of $\abs{\bar{\theta}}$ 
on the $(m_{3/2},M_{\rm CP})$-plane for $y^D = 0.1$ and $m_{\rm soft}=10\,{\rm TeV}$ with $N=3$ (upper panel) and $N=4$ (lower panel) is plotted in Fig.~\ref{fig:summary1}, assuming that the gravitino abundance is equal to the observed DM abundance as in \eq{eq:gravitino_DM}. 
The meshed green and brown colored regions are excluded by the Lyman-$\alpha$ constraint \eqref{eq:Lyman_alpha} and the formation of the domain wall \eqref{eq:T_max_bound}, respectively.
The light blue regions give $\abs{\bar{\theta}}<10^{-10},10^{-11}$. 
The light green shaded region denotes $Y_B=Y_B^{({\rm obs})}$, where $Y_B^{(\rm obs)}$ is the observed baryon-to-entropy ratio. 
Here, $Y_B$ is given by 
Eqs.~\eqref{YbN3} and \eqref{YbN4}, where $m_{\phi}=M_{\rm CP}$ and $\sin (3N\theta_{\rm ini}) \in (0.1,1)$.

The strong CP phase $\abs{\bar{\theta}}$ can be sufficiently small in the light blue region. 
The lower bound on $M_{\rm CP}$ originates from $\bar{\theta}_{\rm radiative}$ 
to avoid a large radiative correction to $\bar{\theta}$ from the heavy vector-like quark.
For a large $m_{3/2}$, the anomaly mediated contribution to the gluino soft mass, $\bar{\theta}_{\rm anomaly}$, becomes large, and hence, there exists an upper bound for $m_{3/2}$.
The upper bound on $M_{\rm CP}$ originates from $\bar{\theta}_{\rm Planck}$ so that the Planck-suppressed CP-violating operator does not reintroduce $\bar{\theta}>10^{-10}$. 
For $N=2$, this bound is too strong, and there is no viable parameter region for our benchmark point.
As the value of $N$ is large, $Z_N$ symmetry strongly forbids the Planck-suppressed operator, and hence, this constraint becomes relaxed for a large $N$, as shown in the figure. 
We can explain the non-observation of the strong CP phase, 
gravitino DM, and baryon asymmetry in the region where the light blue and green shaded regions are overlapped.

\begin{figure}[t]
\begin{center}
\includegraphics[clip, width=11cm]{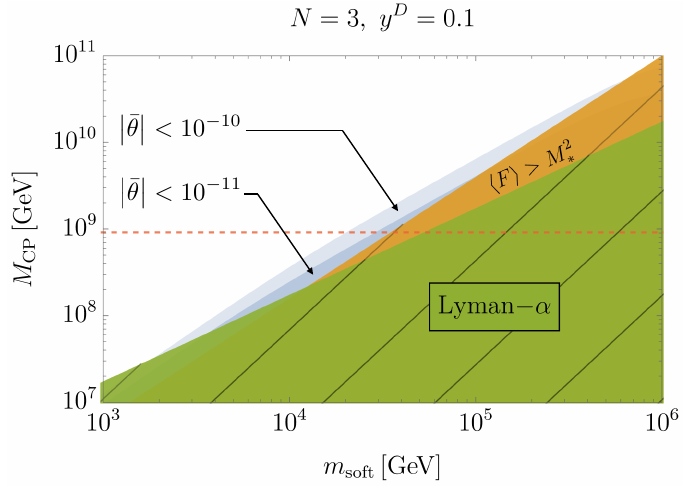}
\\
\vspace{1cm}
\includegraphics[clip, width=11cm]{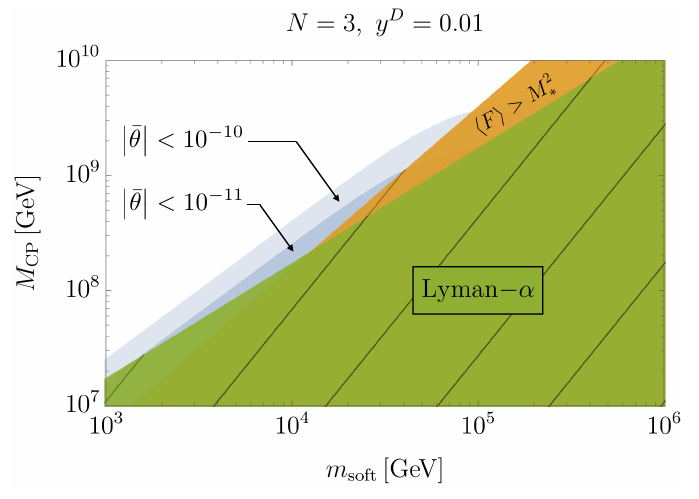}
\end{center}
\caption{Same as Fig.~\ref{fig:summary1} 
but with the observed baryon asymmetry 
on the $(m_{\rm soft},M_{\rm CP})$-plane 
for $y^D=0.1$ (upper panel) and $y^D=0.01$ (lower panel) with $N=3$. 
The meshed green- and brown-colored regions are excluded because of the Lyman-$\alpha$ constraint and by the condition of $\la F \ra < M_*^2$, respectively. 
The red dashed line in the upper panel represents the most conservative lower bound on $M_{\rm CP}$ from the domain wall formation, which can be relaxed for a small value of $H_{\rm inf}$. 
\label{fig:summary2}
}
\end{figure}
Figure~\ref{fig:summary2} shows the prediction of $\abs{\bar{\theta}}$ 
on the $(m_{\rm soft},M_{\rm CP})$-plane 
for $y^D = 0.1$ (upper panel) and $y^D = 0.01$ (lower panel) with $N = 3$. 
Here, $T_{\rm RH}$ and $m_{3/2}$ are taken such that the observed baryon asymmetry and DM energy density are accounted for by the AD mechanism and thermally produced gravitino, respectively. 
The meshed green region is excluded by the Lyman-$\alpha$ constraint. 
The TeV-scale SUSY 
is not compatible with both $\abs{\bar{\theta}}<10^{-10}$ and the Lyman-$\alpha$ constraint 
for the case of $y^D=0.1$ 
but is compatible for a smaller $y^D$. 
The meshed brown region is excluded by the condition of $\la F \ra < M_*^2$. 
For a given $m_{\rm soft}$, 
there is an upper bound on the allowed value of 
$M_{\rm CP}$, which can be relaxed for a large $N$. 
The dashed line in the upper panel represents the lower bound on $M_{\rm CP}$ 
by the domain-wall formation for the case of $H_{\rm inf} \simeq 4.6 \times 10^{13} \GeV$. 
This is the most conservative bound and is relaxed for a small (but large enough to realize the AD mechanism) $H_{\rm inf}$. 
This constraint is evaded in the whole parameter space in the lower panel.

There is a lower bound on $M_{\rm CP}$ to explain the baryon asymmetry for a given $m_{\rm soft}$. 
From Eqs.~\eqref{YbN3} and \eqref{YbN4}, the reheating temperature is related to $M_{\rm CP}$.
Moreover, from \eq{eq:gravitino_DM}, the reheating temperature should be of the same order with $m_{\rm soft}$
to explain DM by the thermally produced gravitino. 
Requiring that $m_{\rm soft}$ is larger than the TeV scale, we obtain 
\beq
 M_{\rm CP} \gtrsim 10^7 \GeV, 
\eeq
for $N = 3,4$, 
where we omit $\mathcal{O}(1)$ factors.

The condition of $\la F \ra < M_*^2$ leads to the upper bound on the SUSY scale,
\beq
 m_{\rm soft} < \frac{\alpha}{4\pi} \sqrt{\sqrt{3} \Mpl m_{3/2}} \, . 
\eeq
If we take the gravitino mass as the lower bound from the Lyman-$\alpha$ constraint, we obtain 
\beq
 m_{\rm soft} < m_{\rm soft}^{(\rm th)} \simeq 1.3 \times 10^4 \GeV. 
\eeq
It should be also noted that the gravitino mass has an upper bound from $\abs{\bar{\theta}} \lesssim 10^{-10}$ because a larger gravitino mass gives a larger contribution from anomaly mediation for any $y^D$ and $N$, as in \eq{eq:theta_anomaly}. 
This implies 
\begin{align}
  &m_{3/2} < \sqrt{3} \Mpl \abs{\bar{\theta}_{\rm anomaly}}^2 \lesssim 40 \MeV, \label{eq:m32 bound amsb}
  \\[1ex]
 &m_{\rm soft} <  \frac{\alpha}{4\pi} \sqrt{3} \Mpl \abs{\bar{\theta}_{\rm anomaly}} 
 \lesssim 1.1 \times 10^6 \GeV ,
\end{align}
where we have used $\abs{\bar{\theta}_{\rm anomaly}} \lesssim 10^{-10}$ in the last inequalities.

\begin{figure}[t]
\begin{center}
\includegraphics[clip, width=11cm]{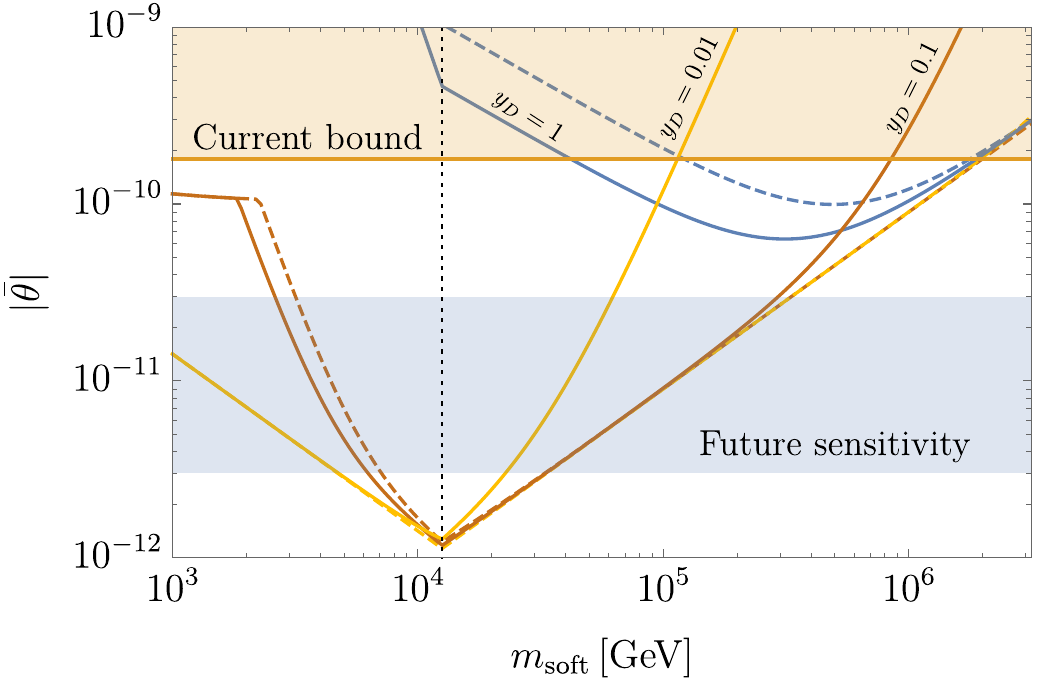}
\end{center}
\caption{
The lower bound on $\abs{\bar{\theta}}$, which is consistent with the baryon asymmetry, gravitino DM, and Lyman-$\alpha$ constraint. 
We take $y^D = 0.01, 0.1$, and $1$ for the yellow, brown and blue lines, respectively. The solid (dashed) lines represent the case of $N=3$ ($N=4$). 
The red-shaded region is excluded from the current constraint on $\bar{\theta}$~\cite{nEDM:2020crw}. The blue shaded region represents the typical sensitivity of future experiments~\cite{nEDM:2019qgk,Ito:2017ywc,Picker:2016ygp,Wurm:2019yfj,Serebrov:2017sqv,n2EDM:2021yah,Chanel:2018zga}.
\label{fig:summary3}
} 
\end{figure}
Eqs.~(\ref{eq:theta_anomaly}) and (\ref{eq:theta_radiative})
indicate that $\bar{\theta}$ is smaller for a smaller $m_{3/2}$. 
Thus, we can obtain the smallest $\bar{\theta}$ for the gravitino mass saturating the Lyman-$\alpha$ bound of \eq{eq:Lyman_alpha}. 
In Fig.~\ref{fig:summary3}, we plot the smallest value of $\bar{\theta}$ as a function of $m_{\rm soft}$, 
by taking the smallest allowed $m_{3/2}$ and determining $M_{\rm CP}$ and $T_{\rm RH}$ to explain the observed baryon asymmetry and DM. 
The value of $m_{3/2}$ is taken to be $5.3 \KeV$ for $m_{\rm soft} < m_{\rm soft}^{(\rm th)}$ 
and $(4\pi/\alpha)^{2} m_{\rm soft}^2/(\sqrt{3} \Mpl)$ for $m_{\rm soft} > m_{\rm soft}^{(\rm th)}$, where $m_{\rm soft}^{(\rm th)}$ ($\simeq 1.3 \times 10^4 \GeV$) is represented by the vertical black dotted line in the figure. 
We take $y^D = 0.01, 0.1$, and $1$ for the yellow, brown, and blue lines, respectively. 
The solid (dashed) lines represent the cases of $N=3$ ($N=4$). 
From the figure, it is evident that there is an absolute lower bound on $\abs{\bar{\theta}}$ ($\sim 10^{-12}$)
even if we change $y^D$ and $N$. This is due to the condition of $\la F \ra < M_*^2$
and the anomaly mediation contribution to $\bar{\theta}$ with a fixed $m_{3/2}$. 
The red shaded region is the current constraint on $\abs{\bar{\theta}}$, where we have used 
$d_n = 10^{-3} \, \bar{\theta}  e \,  {\rm fm}$ (see Refs.~\cite{Vicari:2008jw, Alexandrou:2020mds}
for the uncertainty of this relation)
and the recent upper bound on the neutron EDM, $\abs{d_n} < 1.8 \times 10^{-13}  e \, {\rm fm}$
\cite{nEDM:2020crw} (see also Ref.~\cite{Baker:2006ts}, which yields $\abs{d_n} < 2.9 \times 10^{-13}  e \, {\rm fm}$). 
The blue shaded region represents a typical future prospect for the neutron EDM searches: 
$2\,\text{-}\,3 \times 10^{-15} e \, {\rm fm}$ by nEDM@SNS~\cite{nEDM:2019qgk},
$3 \times 10^{-14} e \, {\rm fm}$ by nEDM search at LANL~\cite{Ito:2017ywc},
$\mathcal{O}(10^{-14}) e \, {\rm fm}$ by TUCAN~\cite{Picker:2016ygp},
$\mathcal{O}(10^{-14}) e \, {\rm fm}$ by PanEDM~\cite{Wurm:2019yfj},
$\mathcal{O}(10^{-(14\,\text{-}\,15)}) e \, {\rm fm}$ by PNPI-ILL-PTI~\cite{Serebrov:2017sqv},
$\mathcal{O}(10^{-(14\,\text{-}\,15)}) e \, {\rm fm}$ by n2EDM~\cite{n2EDM:2021yah}
(see also Ref.~\cite{Chanel:2018zga} for beam EDM experiments). 
Thus, we can obtain information on relatively high scale SUSY, where the soft mass is of the order of $0.1\,\text{-}\,1 \, {\rm PeV}$, depending on $y^D$ and $N$. 
This is an interesting ``smoking-gun'' signal of SUSY in our scenario. 
The prediction of observable $\bar{\theta}$ is unique for the case of spontaneously broken CP symmetry because it is negligibly small for the PQ mechanism.

\section{Conclusions} \label{sec:conclusion}

In the present study, we have investigated the AD baryogenesis scenario in a NB model with gauge-mediated SUSY breaking.
The model introduces a new massive vector-like quark and singlet chiral superfields,
whose scalar components develop VEVs leading to spontaneous breaking of CP symmetry.
The tree-level contribution to $\bar{\theta}$ is forbidden by the structure of the quark mass matrix.
We have performed a detailed analysis of radiative corrections to $\bar{\theta}$ in the NB model with gauge mediation.
We found that the model can be a plausible solution to the strong CP problem.
The AD baryogenesis was realized by using a flat direction including a heavy vector-like quark,
such that long-lived Q-balls do not form after the AD baryogenesis. 
We have shown that the observed baryon asymmetry can be produced in the CP invariant Lagrangian without causing any cosmological problems, such as the domain-wall problem and gravitino overproduction problem. 
The gravitino can be DM and the present DM density can be explained when the reheating temperature is lower than the soft mass scale of the SUSY particles.

The viable parameter region, leading to the successful production of baryon asymmetry and DM, was identified
by several theoretical and observational constraints.
First, the energy scale of spontaneous CP violation is bounded above and below to suppress corrections to the strong CP phase from Planck-suppressed operators and radiative corrections, respectively.
Second, the anomaly-mediated contribution to the gluino mass may violate CP symmetry and should be also suppressed,
which requires a low SUSY breaking scale.
Third, the maximum temperature of the Universe must be lower than the energy scale of spontaneous CP violation
to avoid restoration of CP symmetry.
Otherwise, the domain wall forms after the spontaneous CP violation. 
Finally, the gravitino mass must be sufficiently heavy to evade the Lyman-$\alpha$ constraint on warm DM.

In the viable parameter space, the mass scale of the spontaneous CP-violating sector is larger than $10^7 \GeV$,
the soft mass scale of the visible sector SUSY particles is smaller than $10^6 \GeV$, and 
the gravitino mass is between the Lyman-$\alpha$ bound ($\simeq 5.3 \KeV$)
and the bound from the anomaly-mediation contribution in \eq{eq:m32 bound amsb} ($\simeq 40 \MeV$). 
The strong CP phase originates from Planck-suppressed operators, anomaly mediation, and radiative corrections. 
We have found an interesting prediction of our scenario that
the strong CP phase is larger than $\mathcal{O}(10^{-12})$ and is within the reach of future experiments, 
even if the soft mass scale of the visible sector SUSY particles is much larger than the TeV scale. 
The prediction is unique, particularly compared with the PQ mechanism,
which predicts an extremely small neutron EDM which corresponds to $\bar\theta = {\cal O}(10^{-17})$ \cite{Georgi:1986kr}.

\section*{Acknowledgments}

We would like to thank Jason Evans and Motoo Suzuki for discussions.
The present work is supported by JSPS KAKENHI Grant Numbers
20H05851 (M.Y.), 20K22344 (M.Y.), and 21K13910 (M.Y.), 
World Premier International Research Center Initiative (WPI Initiative), MEXT, Japan.
K.F. is supported by JSPS Grant-in-Aid for Research Fellows Grant No.~20J12415.
Y.N. is supported by Natural Science Foundation of China under grant No.~12150610465.
M.Y. was supported by the Leading Initiative for Excellent Young Researchers, MEXT, Japan.

\bibliography{ref}

\end{document}